\documentclass[useAMS,usenatbib]{mn2e}

\usepackage[T1]{fontenc}
 \usepackage{pslatex}
\usepackage{amsmath}
\usepackage{amsfonts}
\usepackage{color}
\usepackage{url}
\usepackage{graphicx}
\usepackage{subfigure}
\usepackage{amssymb}
\usepackage{soul}
{\newif\ifnotend
\notendtrue
\def\veclist{ABCDEFGHIJKLMNOPQRSTUVWXYZabcdefghijklmnopqrstuvwxyz.}
\def\top#1#2.{#1}
\def\tail#1#2.{#2.}
\loop\expandafter\xdef\csname v\expandafter\top\veclist\endcsname%
{{\noexpand\bf\expandafter\top\veclist}}
\edef\veclist{\expandafter\tail\veclist}
\if\veclist.\notendfalse\fi\ifnotend\repeat}
\def\d{{\rm d}}
\def\e{{\rm e}}

\def\pa{\partial}

\mathchardef\mhyphen="2D

\title[Gas flow in barred potentials]{Gas flow in barred potentials}
\author[Sormani, Binney \& Magorrian]{Mattia C. Sormani$^1$, James Binney$^1$ and John Magorrian$^{1,2}$\\
$^1$ Rudolf Peierls Centre for Theoretical Physics, 1 Keble Road, Oxford
OX1 3NP\\
$^2$ Institut d'Astrophysique de Paris, 98bis boulevard Arago, 75014 Paris}
\begin{document}

\date{}

\def\p{\partial}
\def\Omegap{\Omega_{\rm p}}

\newcommand{\di}{\mathrm{d}}
\newcommand{\bfx}{\mathbf{x}}
\newcommand{\vlos}{\mathrm{v}_{\rm los}}
\newcommand{\Tspin}{T_{\rm s}}
\newcommand{\Tb}{T_{\rm b}}
\newcommand{\degree}{\ensuremath{^\circ}}
\newcommand{\Th}{T_{\rm h}}
\newcommand{\Tc}{T_{\rm c}}
\newcommand{\bfr}{\mathbf{r}}
\newcommand{\bfv}{\mathbf{v}}
\newcommand{\pc}{\,{\rm pc}}
\newcommand{\kpc}{\,{\rm kpc}}
\newcommand{\Myr}{\,{\rm Myr}}
\newcommand{\Gyr}{\,{\rm Gyr}}
\newcommand{\kms}{\,{\rm km\, s^{-1}}}
\newcommand{\de}[2]{\frac{\partial #1}{\partial {#2}}}
\newcommand{\cs}{c_{\rm s}}

\maketitle

\begin{abstract}
We use a Cartesian grid to simulate the flow of gas in a barred Galactic
potential and investigate the effects of varying the sound speed in the gas
and the resolution of the grid. For all sound speeds and resolutions,
streamlines closely follow closed orbits at large and small radii. At
intermediate radii  shocks arise and the streamlines shift between two
families of closed orbits. The point at which the shocks appear and the
streamlines shift between orbit families depends strongly on sound speed and
resolution. For sufficiently large values of these two parameters, the
transfer happens at the cusped orbit as hypothesised by Binney et
al.\ over two decades ago. For sufficiently high resolutions the flow
downstream of the shocks becomes unsteady. If this unsteadiness is physical,
as appears to be the case, it provides a promising explanation for the
asymmetry in the observed distribution of CO.
\end{abstract}

\begin{keywords}
ISM: kinematics and dynamics --
Galaxy: kinematics and dynamics
\end{keywords}

\section{Introduction} \label{sec:introduction}
More than twenty years ago, \cite{binneyetal1991} (hereafter BGSBU) proposed
a consistent picture to explain spectral line emission by different species,
HI, CO and CS, in the Galactic-centre region $| l |<10\degree$ and $|b|<2\degree$. 
They related the flow of gas in an externally imposed, rigidly rotating
barred potential to the structure of the longitude-velocity, $(l,v)$, plane
that one obtains by projecting closed orbits along lines of sight through the disc.
According to BGSBU, gas in the outer parts of the bar drifts slowly towards
the centre while following $x_1$ orbits. These orbits become more and more
elongated as the centre is approached, and eventually the ``cusped orbit'' is
reached, which has a cusp at each end. Interior to the cusped orbit, the
orbits of the $x_1$ family are self-intersecting.  BGSBU hypothesised that
when gas reaches the cusped orbit, it encounters a shock, and then quickly
plunges onto $x_2$ orbits. Thereafter, the gas drifts towards the centre
following $x_2$ orbits.
Using this simple representation of the gas flow, BGSBU provided an appealing
interpretation of observational data: HI emission comes mainly from gas on
non self-intersecting $x_1$ orbits, CO forms as gas is shocked on reaching
the cusped orbit, which explained the characteristic parallelogram-shaped
envelope of CO emission in the $(l,v)$ plane, and CS emission comes from
dense, post-shocked gas flowing on $x_2$ orbits. 

The BGSBU picture has enjoyed considerable success and, alongside the
photometric study of \cite{Blitz1991}, convinced the community that our
Galaxy is barred. In the following years, the presence of the bar has been
confirmed by further photometric evidence
\citep{Stanek1994,Dwek1995,Binney1997}, and there is now little doubt that
the Milky Way is indeed barred.

The BGSBU picture relied on the assumption that gas streamlines follow closed
orbits and that the $x_1\to x_2$ transfer happens at the cusped orbit, and needed validation by hydro simulations.  In support of their
model, BGSBU pointed to simulations by \cite{Athan92b}. However, these
simulations used a different potential from BGSBU, so a natural next step was
to run hydro simulations in the potential BGSBU had used.  One such study
appeared \citep{jenkinsbinney}, but its results were not encouraging: in this
simulation, orbits near the cusped $x_1$ orbit, which played a crucial role
in BGSBU picture, were found to be unoccupied by gas. 

More recently, numerous hydrodynamical simulations have been run with the
goal of understanding the kinematics and dynamics of the cold gas in our
Galaxy \citep{mulderliem,weinersellwood1999,englmaiergerhard1999,
leeetal1999,fux1999,bissantzetal2003,combesrodriguez2008,babaetal2010,Pettitt2014}.
However, none of these simulations use the potential BGSBU used. Some
used a potential inferred from infrared photometric data
\citep{englmaiergerhard1999,bissantzetal2003,combesrodriguez2008}, while
others \citep{fux1999} used the potential generated by a combined hydro and
N-body simulation of the Milky Way. Notwithstanding these efforts, the
interpretation of the observational data remains problematic in some
aspects \citep{SM14}.

Most authors of papers using hydro simulations mention closed orbits, in a
more or less explicit connection with BGSBU, but the literature lacks a
detailed examination of the extent to which good hydro simulations support
the BGSBU picture. The nearest the literature comes to filling this need is
the paper of \cite{jenkinsbinney}, which describes simulations that are of
low resolution by today's standard, and uses sticky particles rather than a
conventional hydro simulation based on the Euler equations. Papers comparing
closed orbits with the results of hydro simulations in bar potentials can be
found \citep{VanAlbadaSanders1982,Athan92a,Athan92b}, but, for lack of
resolution or other reasons, none of them provides sufficient detail to show
which orbits are or are not occupied by gas, especially in the vicinity of
the cusped orbit, which plays a crucial role in the  BGSBU picture. 

In this work, we use high-resolution hydrodynamical simulations to re-examine
the BGSBU picture. In the first part, we test the extent to which the physics of the gas flow hypothesised by BGSBU 
is supported by the simulations. Is the gas flow far from the shocks
well approximated by closed orbits? Can the gas flow be understood as a
transfer from $x_1$ to $x_2$ orbits? Does the transition happen at the cusped
orbit as conjectured by BGSBU? We show that the answers these questions
depend on the spatial resolution and sound speed used in a hydro simulation.
The results are likely to be valid for all barred potentials that have a
general resemblance to the BGSBU potential.  In the second part of the paper,
we discuss the implications of our results for the interpretation of the
observational data. Can we identify in the simulations structures reminiscent
of the CO parallelogram of BGSBU? Under what conditions does the size of the
$x_2$ disc match the region covered by CS emission? Can we explain the high
velocity peaks in the HI $(l,v)$ diagrams at $|v|\simeq270\kms$ and
$|l|\simeq2\degree$?

This paper is structured as follows.  In Sect.~\ref{sec:methods} we present
the numerical schemes employed in the simulations. In Sect.~\ref{sec:results}
we show the results of the hydro simulations.  We discuss the physical
interpretation of the simulations in Sect.~\ref{sec:discussion}, and in
Sect.~\ref{sec:discussion2} we discuss their implications for the
interpretation of observational data. We finally summarise our findings in
Sect.~\ref{sec:conclusion}.
\section{Methods} \label{sec:methods}
\subsection{Hydro Simulation Scheme} \label{sec:hydro}

We assume that the gas is a fluid governed by the Euler equations
complemented by the equation of state of an isothermal ideal gas.  Then we
run two-dimensional hydrodynamical simulations in an externally imposed,
rigidly rotating barred potential.  The output of each simulation consists in
snapshots of the velocity and surface density distributions $\rho(\bfx)$ and
$\bfv(\bfx)$ at chosen times.

We use a grid-based, Eulerian code based on the second-order flux-splitting
scheme developed by \cite{vanAlbada+82} and later used by \cite{Athan92b},
\cite{weinersellwood1999} and others to study gas dynamics in bar potentials.
We used the same implementation of the code as was used by \cite{SM14}, slightly
modified to implement the recycling law of \cite{Athan92b}.
 
This recycling law introduces a term in the continuity equation to take into
account in a simple way the effects of star formation and stellar mass loss.
The equation governing this process is:
\begin{equation}\label{eq:recycle}
\frac{\p \rho}{\p t} =\alpha (\rho_0^2 - \rho^2),
\end{equation}
where $\alpha=0.3 \, M_\odot \pc^{-2}\Gyr^{-1}$ is a constant and
$\rho_0$ is the initial surface density, which is taken to be $\rho_0 = 1 \, M_\odot
\pc^{-2}$.  In practice, the only effect of the recycling law is to prevent
too much gas accumulating in the very centre, and it does not affect the
morphology of the results.  Hence, the results of this paper do not change if
we disable the recycling law. 

We used a grid $N \times N$ to simulate a square $10 \, \kpc$ on a side.  $N$
depends on the resolution of the simulation. For example, if the grid cells
are $\di x= 5\pc$ on a side, we have $N=2000$.  In each run the initial
conditions are as follows.  We start with gas in equilibrium on circular
orbits in an axisymmetrized bar and, to avoid transients, turn on the
non-axisymmetric part of the potential gradually during the first $150\Myr$,
in such a way that the total mass of the underlying potential is conserved in
the process.  We use outflow boundary conditions: gas can freely escape
the simulated region, after which it is lost forever. The potential well is
sufficiently deep, however, that very little gas escapes the regions of
interest.

\subsection{The Potential}
We use the same potential as \cite{jenkinsbinney}.  This arises from two
components.  The first is the bar used by BGSBU, which has the density
distribution:
\begin{equation} \rho_{\rm b}(a) = \rho_{{\rm b}0}
\begin{cases} 
	(a/a_0)^{-\alpha} \qquad \text{if } a\leq a_0 \\
	(a/a_0)^{-\beta}   \qquad \text{if } a > a_0
\end{cases}
\end{equation}
where $a=\sqrt{x^2 + (y^2+z^2)/q^2}$ and the values of the parameters are
$a_0=1.2\kpc$, $\alpha=1.75$, $\beta=3.5$, $\rho_{{\rm b}0}=0.69 M_\odot \pc^{-3}$,
$q=0.75$, so  the major axis of the bar always lies along the $x$ axis.
The second component is a razor-thin exponential disc, which
has been added to complete realistically the circular velocity curve outside
$R\simeq1\kpc$, and it has little influence inside this radius. The exponential
disc is generated by a surface density distribution
\begin{equation} 
\Sigma(R) = \Sigma_0 \e^{-R/R_{\rm d}},
\end{equation} 
where $R$ is the radius in cylindrical coordinates and the parameters have values $\Sigma_0
= 1300\,M_\odot \pc^{-2}$, $R_d = 4.5 \kpc$.

The potential is assumed to be rigidly rotating with constant pattern speed $\Omega_{\rm p} = 63 \kms \kpc^{-1}$. This places the Inner Lindblad Resonance at $R_{\rm ILR} = 0.6\kpc$ and corotation at $R_{\rm CR} = 3.7\kpc$.

\subsection{Projecting to the $(l,v)$ plane} \label{sec:projection}
 We adopt a very simple projection procedure to produce the predicted $(l,v)$
distributions for each simulation snapshot
($\rho(\bfx),\bfv(\bfx)$). Throughout this paper, we assume that the Sun is
undergoing circular motion at a radius $R_0=8\kpc$ with speed $v_\odot =
220\kms$.  
Calling $\phi$ the angle between the major axis and the Sun--GC line, the
Cartesian coordinates of the Sun are given by $x_\odot = R_0 \cos \phi$,
$y_\odot = R_0 \sin \phi$. All the projections made in this paper assume
$\phi=20\degree$.

The resolution of our $(l,v)$ diagrams is $\Delta l=0.25\degree$ in longitude
and $\Delta v=2.5\kms$ in velocity.  Along each line of sight, we sample the
density and the velocity by linearly interpolating the results of the
simulations at points separated by $\delta s =1 \pc$.  These density measures
are accumulated in velocity bins of width $\Delta v=2.5\kms$.  The final
$(l,v)$ intensity at the chosen longitude in each range of velocity is the
mass in the relevant bin divided by the square of its distance. 

This procedure yields a predicted brightness temperature that is linear in
column density so it is equivalent to the simplest radiative transfer
calculation.  In the case of HI, the brightness temperature is linear in the
column density if the gas has constant spin temperature and its optical depth
is negligible.  So our projection is equivalent to simple HI radiative
transfer in the constant-temperature, optically-thin case.  The assumption of
constant temperature is known to be a simplification for Galactic HI, which
is instead often modelled as a medium made by two or more phases at different
temperatures \citep[see for example][]{Ferriere2001}.  In the case of
${}^{12}$CO, the brightness temperature is not linearly related to density
when considering a single cloud, but a linear relationship will hold between
brightness temperature and the number density of unresolved CO clouds
provided the cloud density is low enough for shadowing of clouds to be
unimportant \citep[see, e.g.,][\S 8.1.4]{BM}.

\section{Results} \label{sec:results}
\begin{figure}
\includegraphics[width=0.46\textwidth]{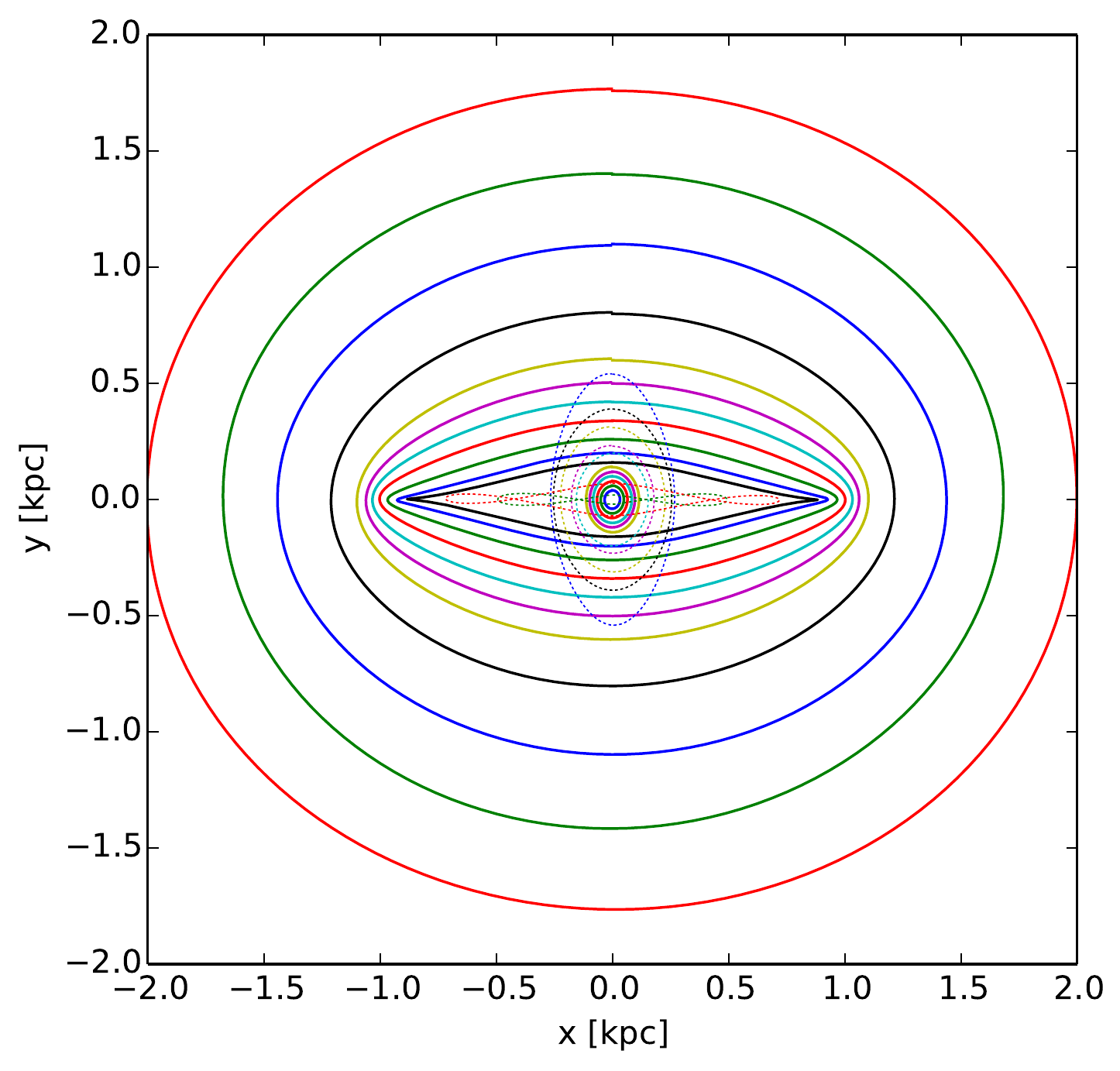} \caption{$x_1$
and $x_2$ orbits in the potential used by BGSBU. Orbits are shown  in the
frame that rotates (clockwise) with the bar. The horizontally elongated
 orbits form the $x_1$ family, while the vertically elongated ones are from
the $x_2$ family.  BGSBU hypothesised that the orbits drawn in full lines
have gas, while orbits those shown dashed are unoccupied. The cusped $x_1$
orbit is the smallest horizontally elongated orbit drawn in full lines and is
shown black.} \label{fig:orb1}
\end{figure}

Fig.~\ref{fig:orb1} shows a selection of closed orbits in the BGSBU
potential, in the frame corotating with the bar. Orbits that BGSBU believed
to carry gas are shown by full lines, while those they thought empty are
shown by dashed lines. In the outer region we show a nested sequence of non
self-intersecting $x_1$ orbits that terminates in the cusped orbit, drawn in
black. Inside this orbit we show dotted two self-intersecting $x_1$ orbits.
The vertically elongated orbits belong to the $x_2$ family, which extends quite a bit beyond the point where the cusped $x_1$ orbit intercepts the vertical axis. 
BGSBU argued that gas transfers between the $x_1$ and $x_2$ families at the cusped
orbit. A primary goal of this paper is to test this conjecture. 

\begin{figure*}
\includegraphics[width=1.0\textwidth]{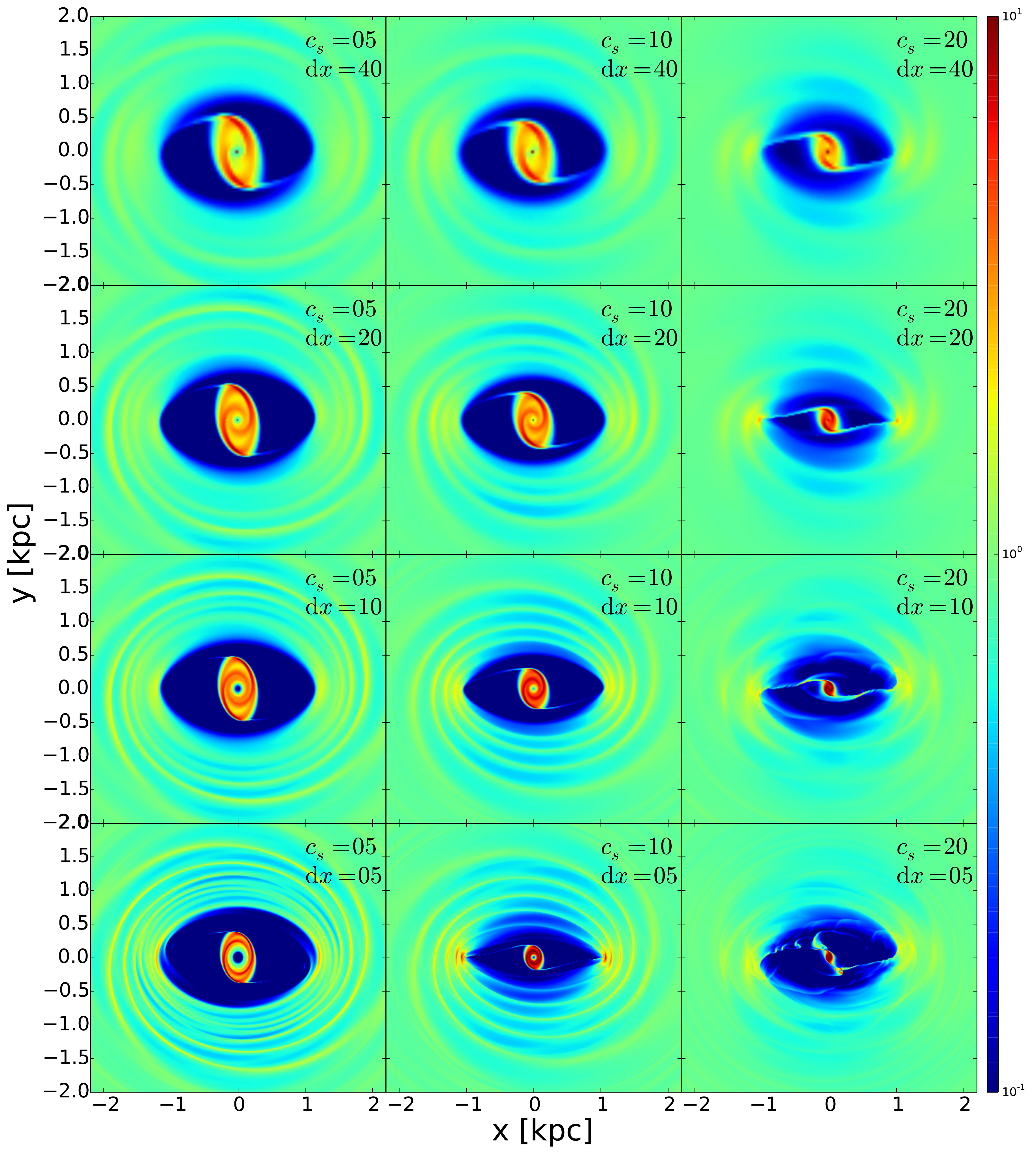}
 \caption{The fluid density in hydro simulations in the BGSBU potential for
different spatial resolutions and sound speeds. $\cs$ is increasing left to
right taking values 5,10,$20\kms$.  $\di x$ is decreasing from top to bottom
taking values 40,20,10,$5\pc$. Gas has reached an approximately steady state in
the rotating frame and circulates clockwise. All snapshots are taken at
$t=280\Myr$.} \label{fig:res1}
\end{figure*}

Fig.~\ref{fig:res1} summarises the results of our simulations.  It
shows the density of hydro simulations for different grid spacings $\di
x$ and sound speeds $\cs$. All snapshots are taken at the same
time $t=280\Myr$. Some common features of the gas flow can be identified in
all panels. In the outer part, approximately corresponding to the green
region, the gas follows $x_1$ orbits. At some point near the $x$
axis, two thin offset shocks emerge, which connect the green region to the
reddish central disc. The central disc is called the $x_2$ disc and
is made by gas on $x_2$ orbits. Most of the gas plunges to the $x_2$ disc or
``central molecular zone''
through the shocks.

Thus, the gas follows the $x_1$ orbits in the outer part and the $x_2$ orbits
in the central part, with a transition zone containing the shocks in
between. In each simulation, we can identify an innermost occupied $x_1$
orbit. The shocks are formed just after this orbit and they induce the
transition from the $x_1$ to the $x_2$ family. We call this innermost
occupied $x_1$ orbit the transition orbit, and the transition point its
position in a parametrisation of the sequence of $x_1$ orbits. 

\begin{figure*}
\includegraphics[width=1.0\textwidth]{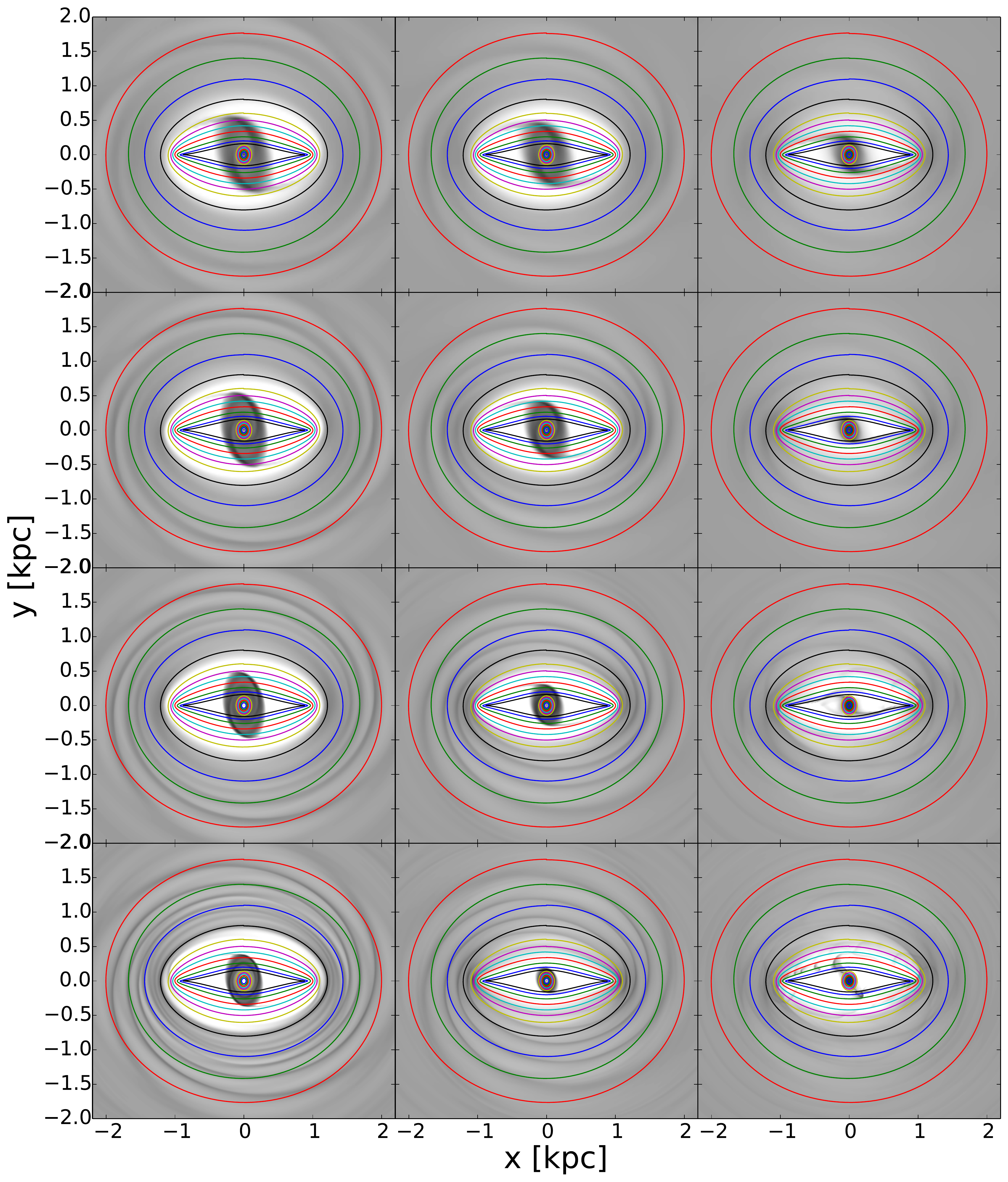}
\caption{Same as Fig.~\ref{fig:res1} but with superimposed orbits.}
\label{fig:res2}
\end{figure*}

BGSBU assumed that the transition orbit is the cusped orbit. In our
simulations, the transition point and the size of the $x_2$ disc depend
strongly on both the resolution and the sound speed.  Consider, for example,
the middle column in Fig.~\ref{fig:res1}, corresponding to $\cs=10\kms$.  As
we increase the resolution, the transition point moves inwards while the
$x_2$ disc shrinks.  At the highest resolution, $\di x=5\pc$, the transition
orbit almost coincides with the cusped orbit, as predicted by BGSBU.  At
lower resolution the transition happens earlier and the transition orbit is much
bigger than the cusped orbit.  Fig.~\ref{fig:res2} shows the same density
snapshots as Fig.~\ref{fig:res1}, superimposed on the closed orbits that
BGSBU thought carried gas for a better comparison.

Increasing the sound speed also has the effect of postponing the transition
and shrinking the $x_2$ disc.  Consider, for example, the second row in
Fig.~\ref{fig:res1}, corresponding to $\di x=20\pc$.  At this resolution, the
transition happens very early, for $\cs=5\kms$, and the transition orbit is
quite an outer $x_1$ orbit (approximately the yellow orbit in
Fig.~\ref{fig:orb1}).  As we increase the sound speed, the transition orbit
moves inwards, and for $\cs=20\kms$ it coincides in Fig.~\ref{fig:orb1} with
the green orbit that lies just outside the cusped orbit.  We will discuss the
origins of these systematics in
Sect.~\ref{sec:discussion}. 

\begin{figure*}
\includegraphics[width=0.98\textwidth]{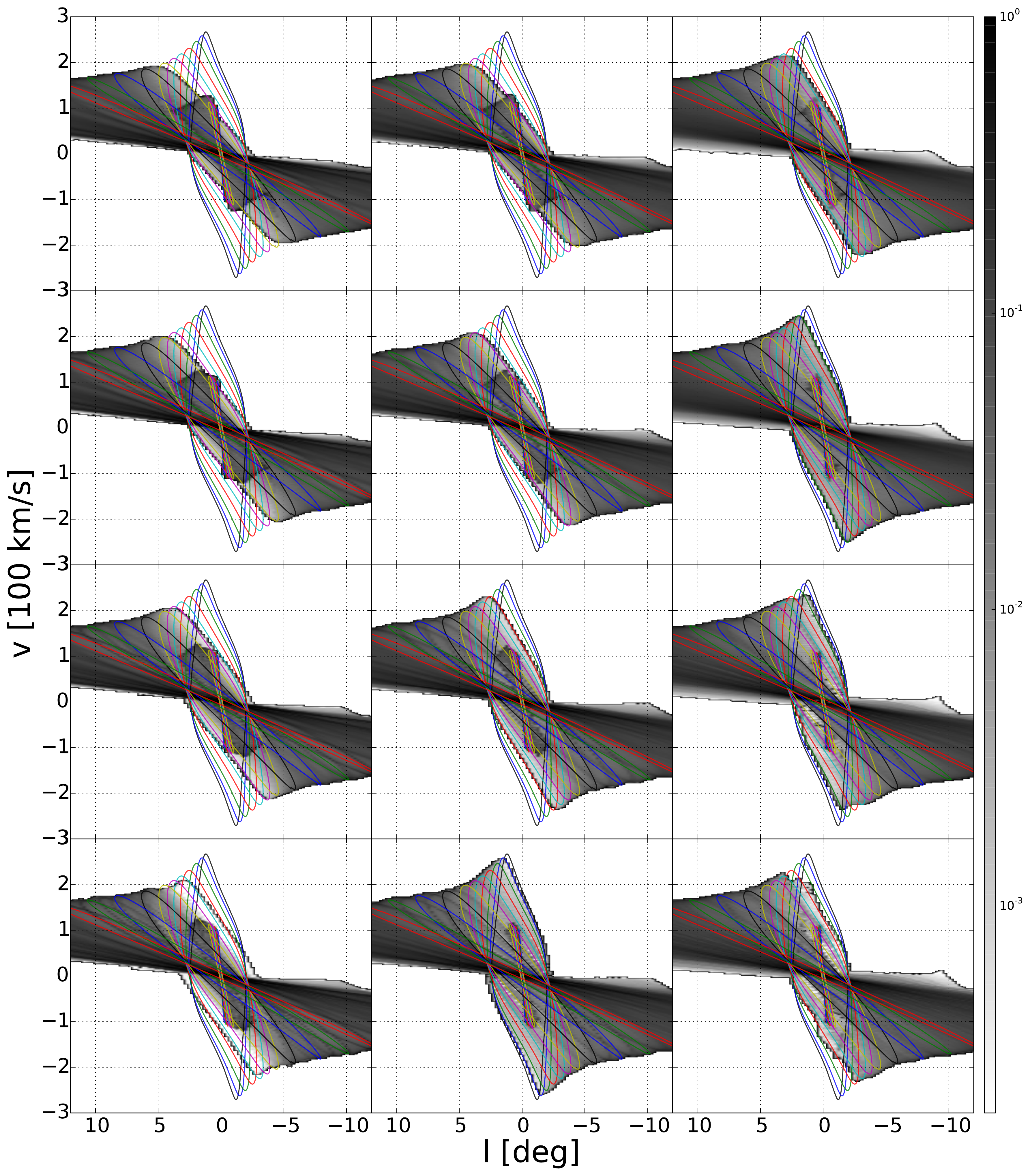}
 \caption{The simulations of Figs.~\ref{fig:res1} and \ref{fig:res2}
projected into the $(l,v)$ plane.  Solid lines show the $(l,v)$ traces of closed
orbits, with colours matching other figures.  The Sun is assumed to be in a
circular orbit with $v=220\kms$ at $R_0=8\kpc$, and the bar major axis makes
an angle $\phi=20\degree$ with the Sun-Galactic centre line.}
\label{fig:res3}
\end{figure*}

\begin{figure*}
\includegraphics[width=1.0\textwidth]{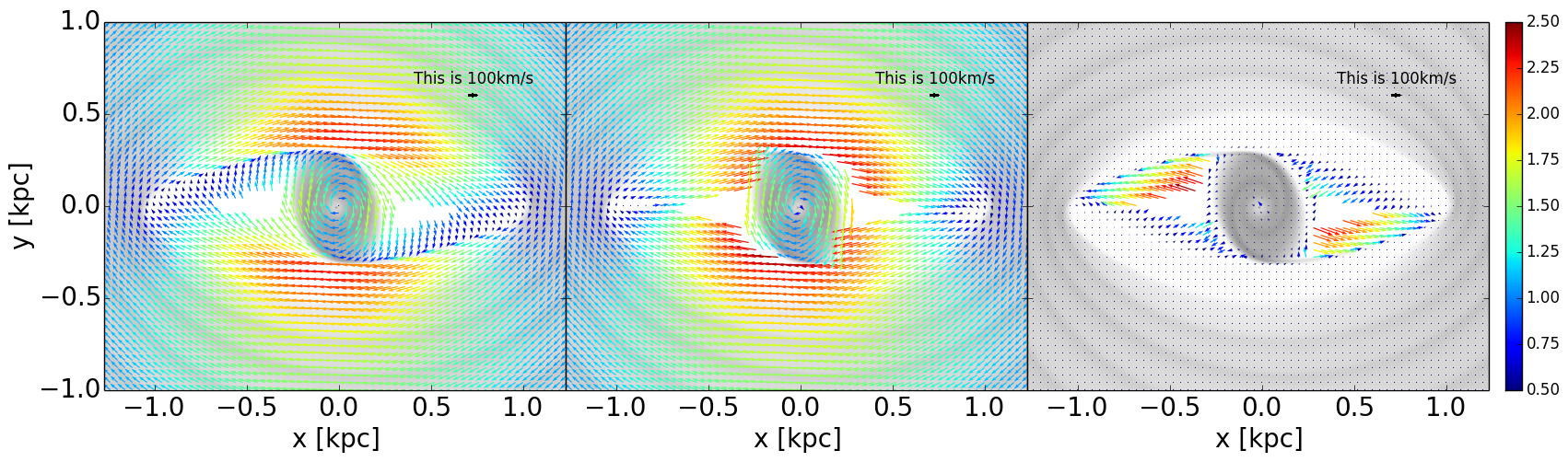}
 \caption{Using closed orbits to approximate the hydro velocity field for the
simulation with $\di x=10 \pc$, $\cs = 10 \kms$ shown at left. The central
panel shows the best approximation to that velocity field obtained using
closed orbits belonging to $x_1$ or $x_2$ family. For each point where more
than one orbit is found, the chosen velocity is the one closest to the hydro
velocity. The right panel shows the vector differences between the
left and middle panels. In all panels points without  orbits are
masked and the density of the hydro simulation is visible in the background. The
colorbars show the speed at each point in units of $100 \kms$.}
\label{fig:vfields}
\end{figure*}

The two highest-resolution simulations in the right column of
Fig.~\ref{fig:res1} look peculiar.  These correspond to $\cs=20\kms$ and $\di
x = 5,10 \pc$.  At the time shown, all other simulations have already reached
an approximate steady state, and they would not appear significantly different after another $250\Myr$ or more.
These two simulations instead manifest a complex
unsteady flow interior to the transition orbit.  The shocks are not stable,
but keep forming and dissolving in an endless cycle.  In some snapshots, they
are almost completely formed and smooth (see also Sect.
\ref{sec:instability} and Fig.~\ref{fig:instability2}). At these moments,
they lie very close to the $x$ axis. A little later, vortices grow on the
leading side, which move around and eventually bump on the opposite shock,
creating more vorticity. Later the cycle repeats. Tests described in
Sect.~\ref{sec:instability} suggest that the unsteadiness is real and not an
artifact of our particular code.

So far we have not discussed the velocity structure of our simulations, but
this is crucial for the interpretation of observations. Fig.~\ref{fig:res3}
shows the projections into the $(l,v)$ plane of the snapshots of
Fig.~\ref{fig:res1} for an assumed angle $\phi=20\degree$ between the bar's
major axis and the Sun-Galactic centre line. The lines in Fig.~\ref{fig:res3}
show the $(l,v)$ traces of some of the closed orbits plotted in
Fig.~\ref{fig:res2} using the same colour scheme -- we plot only the orbits
that BGSBU thought occupied (those shown with full lines in
Fig.~\ref{fig:orb1}).

For all resolutions and sound speeds, in Fig.~\ref{fig:res3} the envelope of the
outermost $x_1$ orbits matches the envelope of the hydro distribution very
well.  As we move towards smaller values of $|l|$, the traces of orbits sweep
up towards the high-velocity peak of the cusped orbit, but at the transition
orbit the hydro envelope starts to fall, and thus becomes separated from the
orbit envelope.  The projection of the hydro $x_2$ disc is clearly
identifiable as a darker region near the centre. At the smallest longitudes
its boundary is delineated by the traces of the $x_2$ orbits, but for low
sound speeds or resolutions it extends far beyond the region covered by the
plotted
$x_2$ orbits.

This finding again confirms the picture of the gas flow that we delineated
above.  When the transition point is too early, the innermost non-self
intersecting $x_1$ orbits, which are populated by gas in the BGSBU picture,
are void of gas in the hydro simulations. Outer $x_2$ orbits that lack gas in
the BGSBU picture are occupied by gas in the hydro simulation.  As we
increase the resolution at $\cs=10\kms$ (middle column), the transition
between the $x_1$ and $x_2$ families moves inwards, and the hydro envelope
matches more and more closely the predictions of the BGSBU picture. At the
same time, the projection of the $x_2$ disc shrinks, also approaching the
BGSBU picture.  Eventually, for $\di x = 5\pc$ and $\cs=10\kms$, the envelope
of the hydro the projection of the $x_2$ disc  match very well as in the
BGSBU picture.

At low sound speed (left column of Fig.~\ref{fig:res3}), the transition always happens early, so
orbits close to the cusped orbit are unoccupied at all resolutions.
Therefore at low sound speeds the hydro simulations gives results
inconsistent with the BGSBU picture.  At high sound speeds the hydro
simulations approach the BGSBU picture as we increase the resolution at
first.  But at higher resolution, unsteadiness  makes the projected hydro
deviate significantly from the BGSBU picture.

To explain this situation from a face-on perspective, we take as an
illustrative example the case $\di x=10$, $\cs=10\kms$.
Fig.~\ref{fig:vfields} compares the velocity field of the hydro simulation
with that of the closed orbits. The left panel shows the velocity field of
the hydro simulation. The central panel shows the best approximation to the
hydro velocity field that can be obtained from the $x_1$ and $x_2$ orbits: at
each point where more than an orbit is present, we show the velocity of the
orbit that best matches the hydro velocity field. In all panels the density of the hydro simulation
is shown in the background.  Locations through which no closed orbit
passes are masked out. The right panel is the most interesting panel:
it
shows the vector difference between the left and middle panels, and shows
clearly that the outer $x_1$ orbits and the inner $x_2$ orbits both reproduce
the hydro velocity field accurately. The red orbit in Fig. \ref{fig:orb1} is the transition orbit in this case,
the last orbit at which the velocity fields coincide, and is just outside the
shocks. After this, the shocks emerge and the two velocity fields suddenly
diverge.

At different resolutions and sound speeds, the situation is qualitatively
very similar but the point of transition between orbit families changes.  For
$\cs=20\kms$ and high resolution the hydro flow in the transition region is
unsteady.

\section{The physics of  the gas flow} \label{sec:discussion}

\subsection{dependence on the sound speed} \label{sec:cs}
In Sect. \ref{sec:results}, we have seen that as the sound speed increases,
the transition point and the shocks move inwards.  At the same
time, the shocks also become more horizontal and closer to the $x$ axis.
\cite{englmaiergerhard1997} and \cite{PatsisAthan2000} found similar
results when varying the sound speed.  A qualitative explanation for this
behaviour is as follows \citep[for another discussion see
also][]{englmaiergerhard1997}. 

The more elongated an $x_1$ orbit is, the larger the variation along the
orbit in the speed of an orbiting particle. Since the speed decreases as
particles move from the orbit's minor axis to its major axis, the density of
gas that is streaming along the orbit increases as the major axis is
approached. When the sound speed is high, pressure assists gravity in slowing
gas as it approaches the major axis, and equally assists gravity in
accelerating the flow after the major axis has been passed. In the absence of
a shock, the gas is reversibly compressed and decompressed so it can continue
to keep close to one orbit for several revolutions.

If the sound speed is too low, the convergence of the gas as the major axis
is approached leads to shock formation.  Entropy is created in
the shock, so the decompression after the major axis has been passed does not
reverse what happened as the axis was approached, and the flow deviates
strongly from orbits.

The key to avoiding shock formation is the ability of sound waves to
carry information about fluctuations in density upstream so oncoming gas can
be slowed in a timely manner when the density increases at a
downstream location. The nearer the cusped orbit is approached, the larger is
the velocity gradient up which sound waves have to travel if a shock is to be
avoided. Hence decreasing the sound speed causes the shock to form further
out. 

\subsection{Dependence on numerical resolution}\label{sec:dx}

\begin{figure*}
\includegraphics[width=0.9\textwidth]{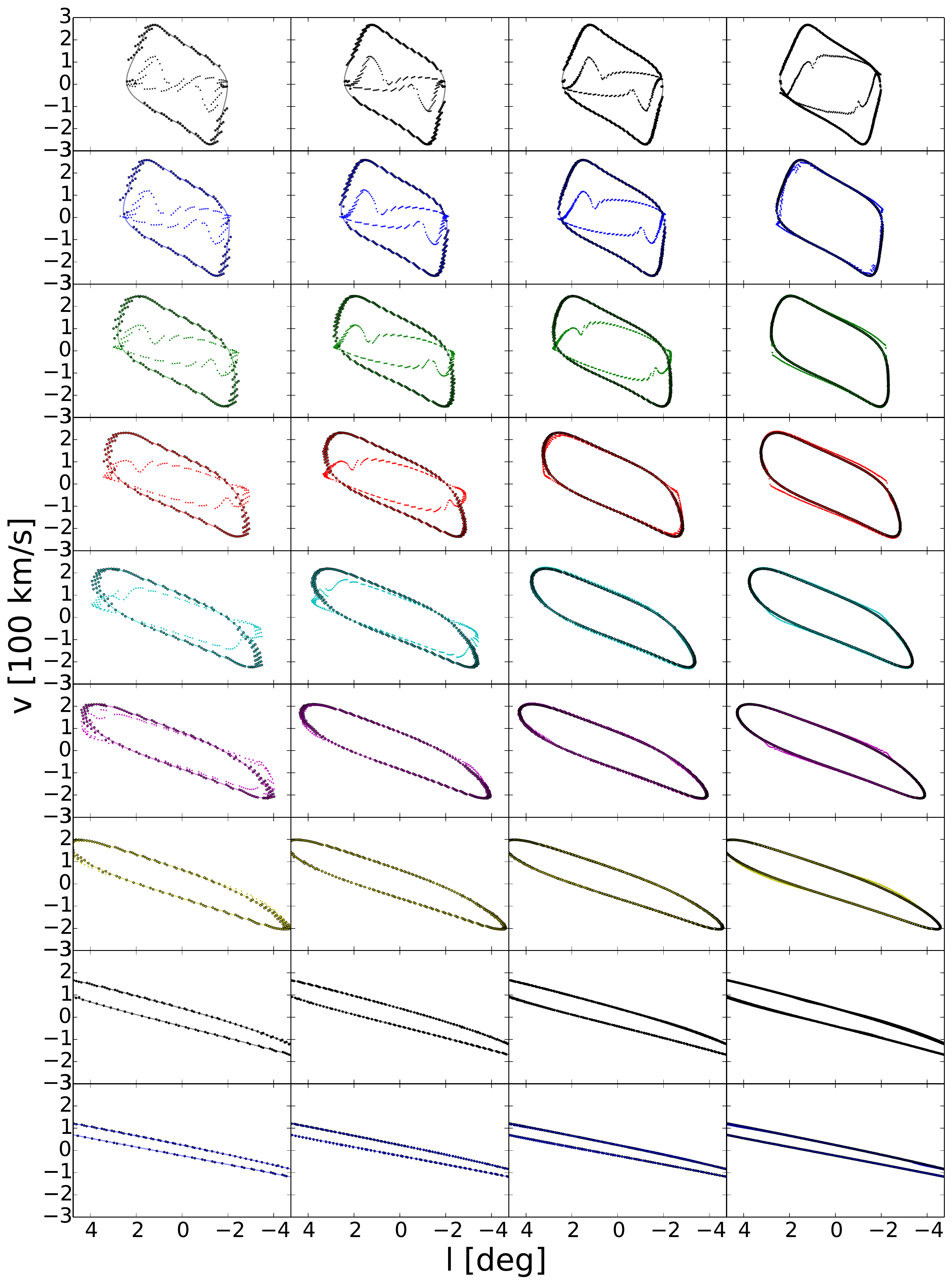}
 \caption{Each panel refers to a particular $x_1$ orbit. In full lines, the
trace of the orbit. The bigger, empty dots are the velocity field of closed
$x_1$ orbits sampled with the same resolution of a hydro simulation along the
$xy$ trajectory of the orbit. Smaller, filled dots are the projection of a
hydro simulation cells along the same trajectory. Each column refers to one
hydro simulations. From left to right, the resolution increases from $\di
x=40 \pc$ to $\di x = 5 \pc$. All simulations are for $\cs=10\kms$.}
\label{fig:res4}
\end{figure*}
Increasing the resolution moves the orbit at which shocks form inwards with
important consequences for the interpretation of the gas flow in our Galaxy
that will be discussed in Sect.~\ref{sec:discussion2}. A likely explanation
for this phenomenon is that at low resolution the innermost $x_1$ orbits are
inadequately resolved. To resolve the cusp we need in principle infinite
resolution.  By increasing the resolution, we can resolve orbits closer and
closer to the cusped one, and gas can settle on these orbits, delaying the
formation of shocks.

Fig \ref{fig:res4} tests this idea by showing the $(l,v)$ traces of nine
orbits from the cusped orbit (at the top) outwards. Each orbit is mapped four
times at resolutions that increase from left to right. The faint solid lines
show the traces of the orbits and as such are the same along every row. The
big points scattered around this line show the $(l,v)$ trace one obtains by
associating each hydro cell through which the orbit passes with the mean
velocity of its passage, and then projecting the resulting velocities of the
visited cell. The small points show the projection at
the visited cells of the hydro velocity field for $\cs=10\kms$.

We see that orbits that are well outside  the cusped orbit (lower part of
the figure) are well represented by even the coarsest grid (left side of the
figure), and, moreover, their velocities are accurately reproduced by the
hydro code. As we move up the figure and therefore approach the cusped orbit,
at low resolutions the big dots scatter widely around the orbit's curve. The
scatter is widest at the largest values of $|l|$ because that is where the
velocity gradient is largest and thus the finite resolution of the grid has its
biggest impact. For each orbit outside the cusped orbit there is a resolution
finer than which the hydro velocity field is almost the same as the orbital
velocity because along that orbit the flow does not encounter a shock.

\subsection{Unsteady flow} \label{sec:instability}
As mentioned in Sect. \ref{sec:results}, the flow becomes unsteady at
high sound speed and high resolution. To test whether this phenomenon
was an artifact generated by our code, we re-ran some simulations with
a completely different code, PLUTO \citep{Pluto2007}.
This is free software for the numerical solution of systems of
conservation laws targeting high Mach number flows in astrophysical
fluid dynamics.
It has a modular structure that makes it easy to change the algorithm
used to simulate a flow.
We use this modularity to investigate the effects of the choice of
Riemann solver and flux limiter.
In all of our runs we used a static Cartesian mesh and RK2
time-stepping.
The code was modified to implement our recycling law
\eqref{eq:recycle}.

\begin{figure}
\includegraphics[width=0.5\textwidth]{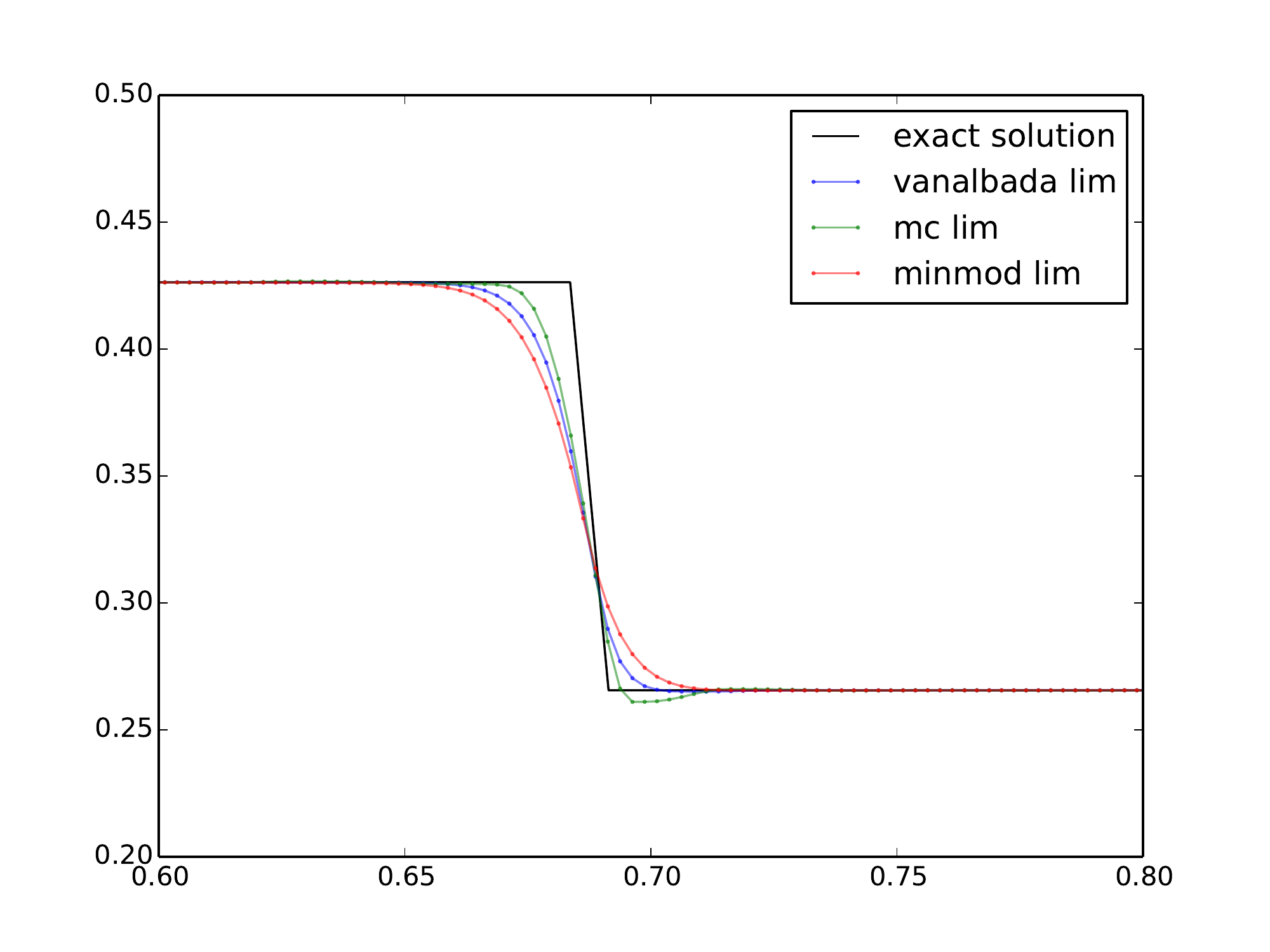}
\caption{Results of SOD shock tube test problem at $t=0.2$ for different flux limiters. This figure zooms around a shock. The resolution is $\di x=0.0025$.}
\label{fig:sod}
\end{figure}

\begin{figure}
\includegraphics[width=0.5\textwidth]{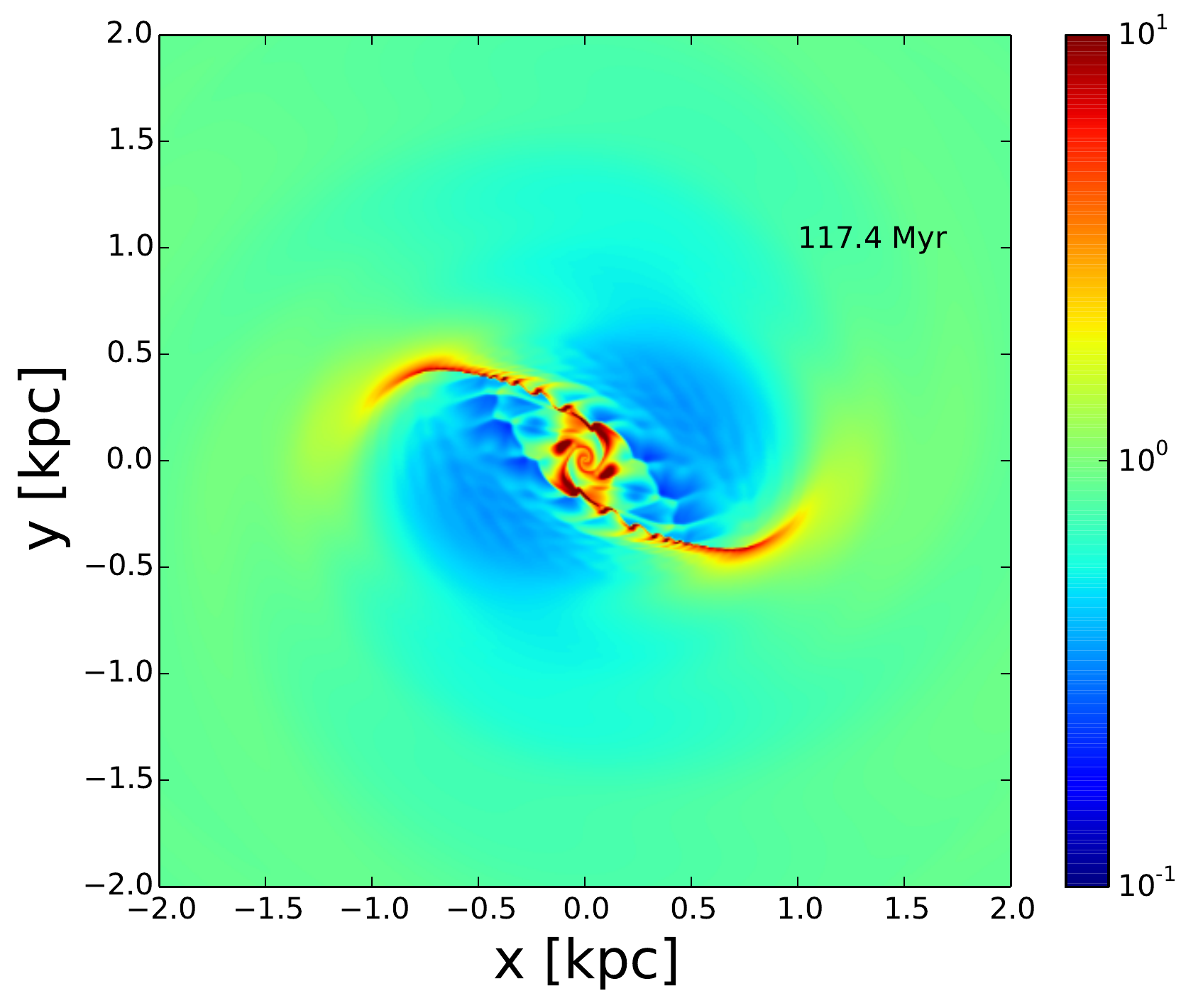}
\includegraphics[width=0.5\textwidth]{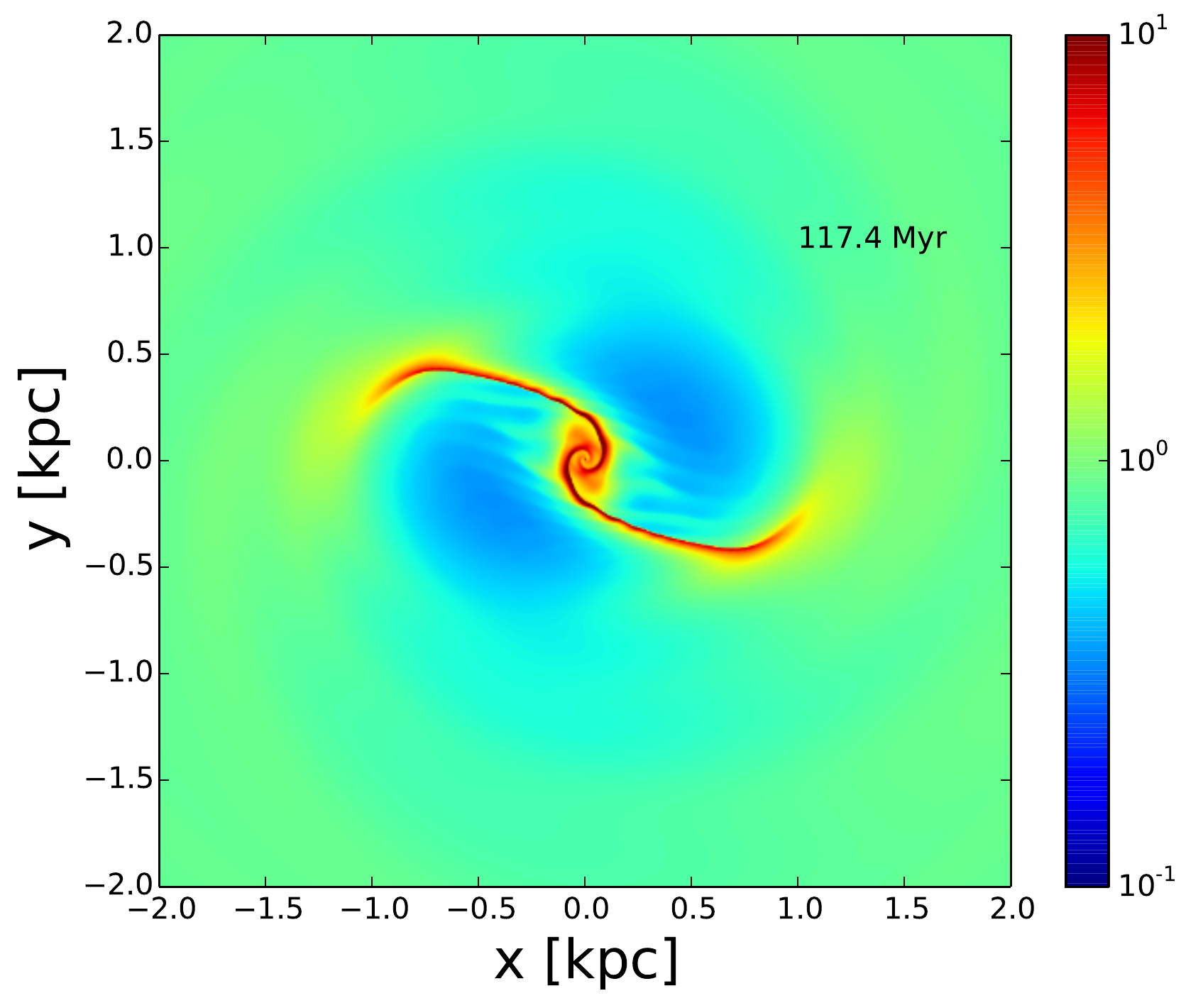}
\includegraphics[width=0.5\textwidth]{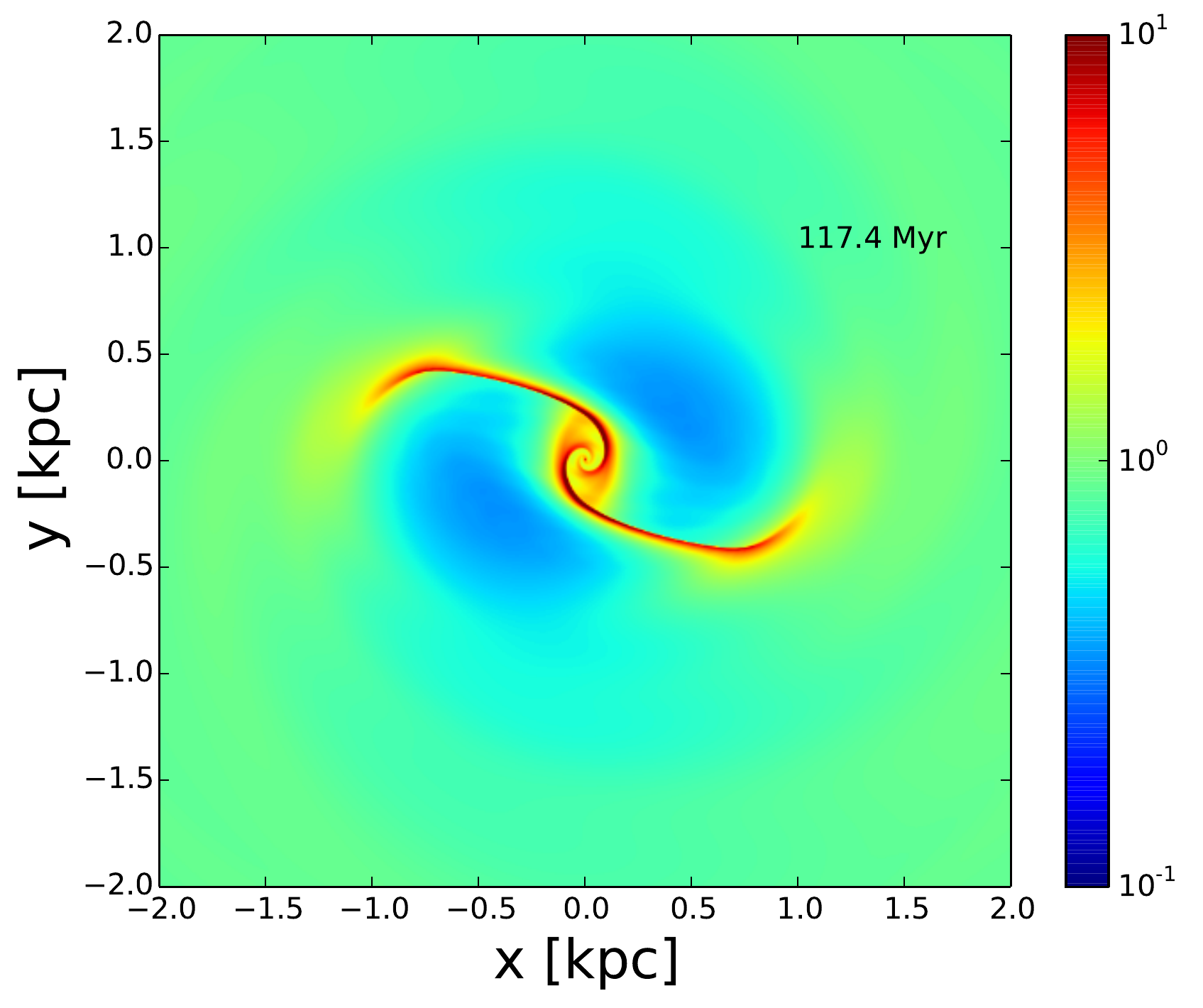}
\caption{Three snapshots capturing the moment when the instability is forming. The three snapshots come from three different simulations obtained with the Pluto code. The only 
difference between the three is the limiter used. On top using the MC limiter, middle the VanAlbada limiter and bottom the MidMod limiter.}
\label{fig:instability}
\end{figure}

We found that the Roe and HLL Riemann solvers produced the same
results as each other, and our own code.
The choice of flux limiter, by contrast, does
have a significant impact on the computed flow: unsteadiness of the flow can
be suppressed by changing to a more diffusive flux limiter.  Flux limiters
are used in numerical schemes to handle the flow close to discontinuities
such as shocks. We tested three different limiters and found that
unsteadiness is suppressed by choosing a more diffusive limiter, which has a
higher numerical viscosity and produces thicker shocks. Fig.~\ref{fig:sod}
shows the result of using PLUTO to perform a standard one-dimensional test
problem, the SOD shock tube \citep{Sod78}. It shows that the three limiters
produce shocks of different thicknesses. The most diffusive (more viscosity)
is the MINMOD limiter \citep{Roe1986}. The Van Albada limiter used in this
paper \citep{vanAlbada+82} has intermediate diffusivity, while the least
diffusive limiter (least viscosity) is the MC limiter \citep{VanLeer1977}.
Fig.~\ref{fig:instability} shows a snapshot at an intermediate time for
simulations with $\di x=5\pc$, $\cs=20\kms$ obtained using PLUTO. The only
difference between the three simulations is the limiter used. From top to
bottom the figure shows the flows obtained with limiters of increasing
diffusivity. The top flow has much irregular and unsteady structure that is
completely absent from the bottom flow. The central panel shows an
intermediate level of unsteady structure.

In these flows unsteady features seem to arise at the shocks, which develop
wrinkles  that shed vortices. The vortices then move away from the shocks.
Vorticity is generated at the shocks and propagates away from them into the
body of the flow.  

\begin{figure}
\includegraphics[width=0.5\textwidth]{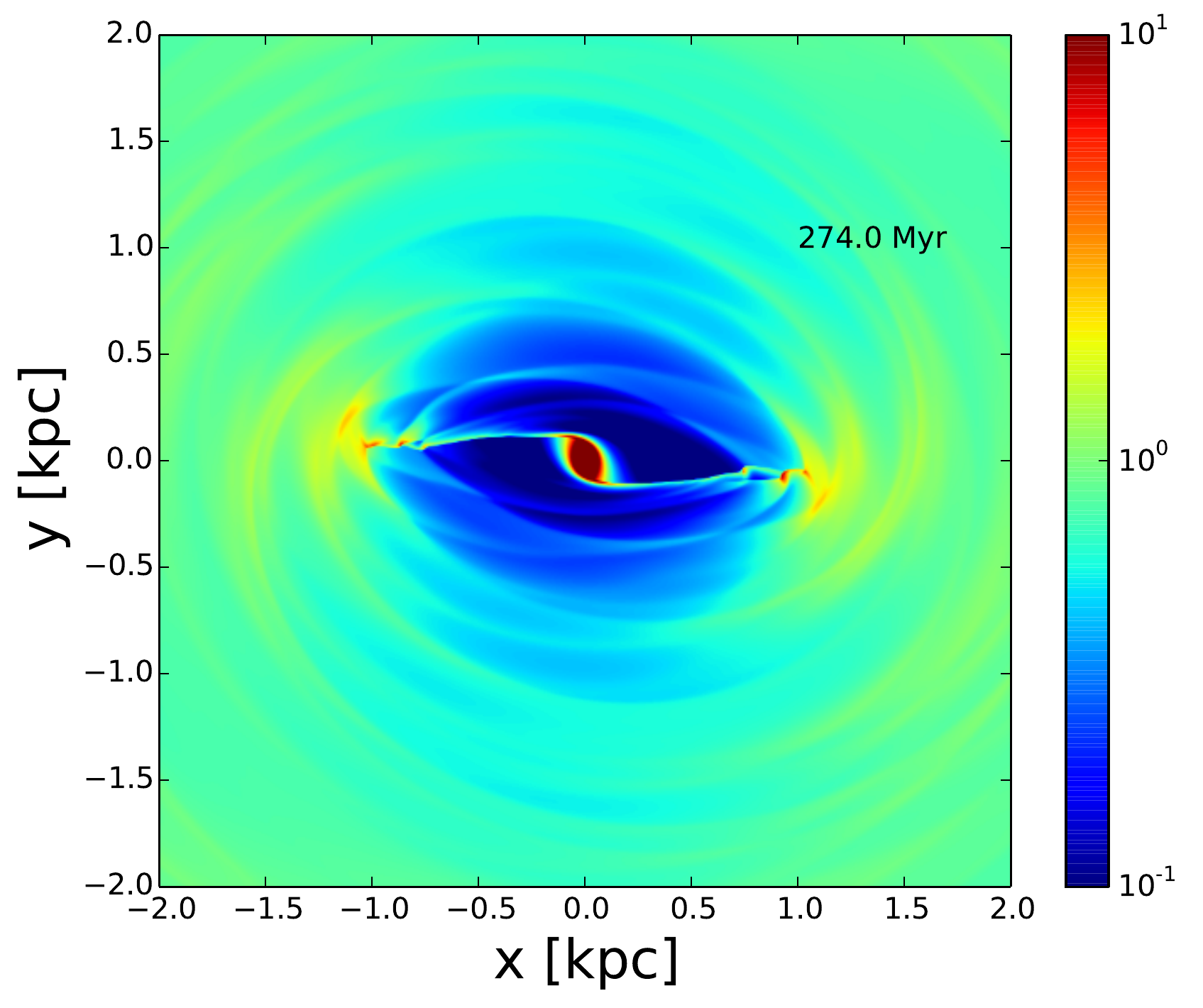}
\includegraphics[width=0.5\textwidth]{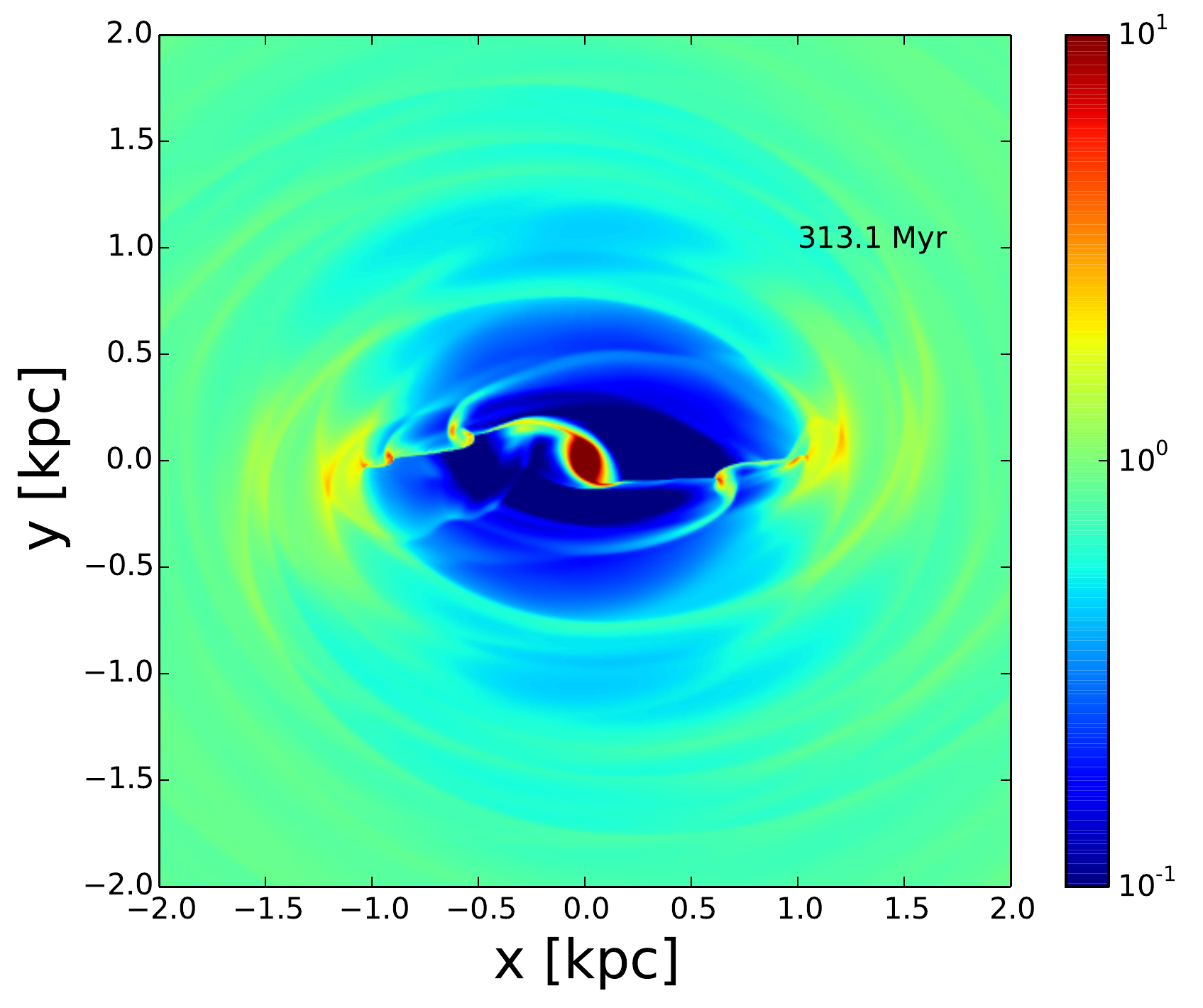}
\includegraphics[width=0.5\textwidth]{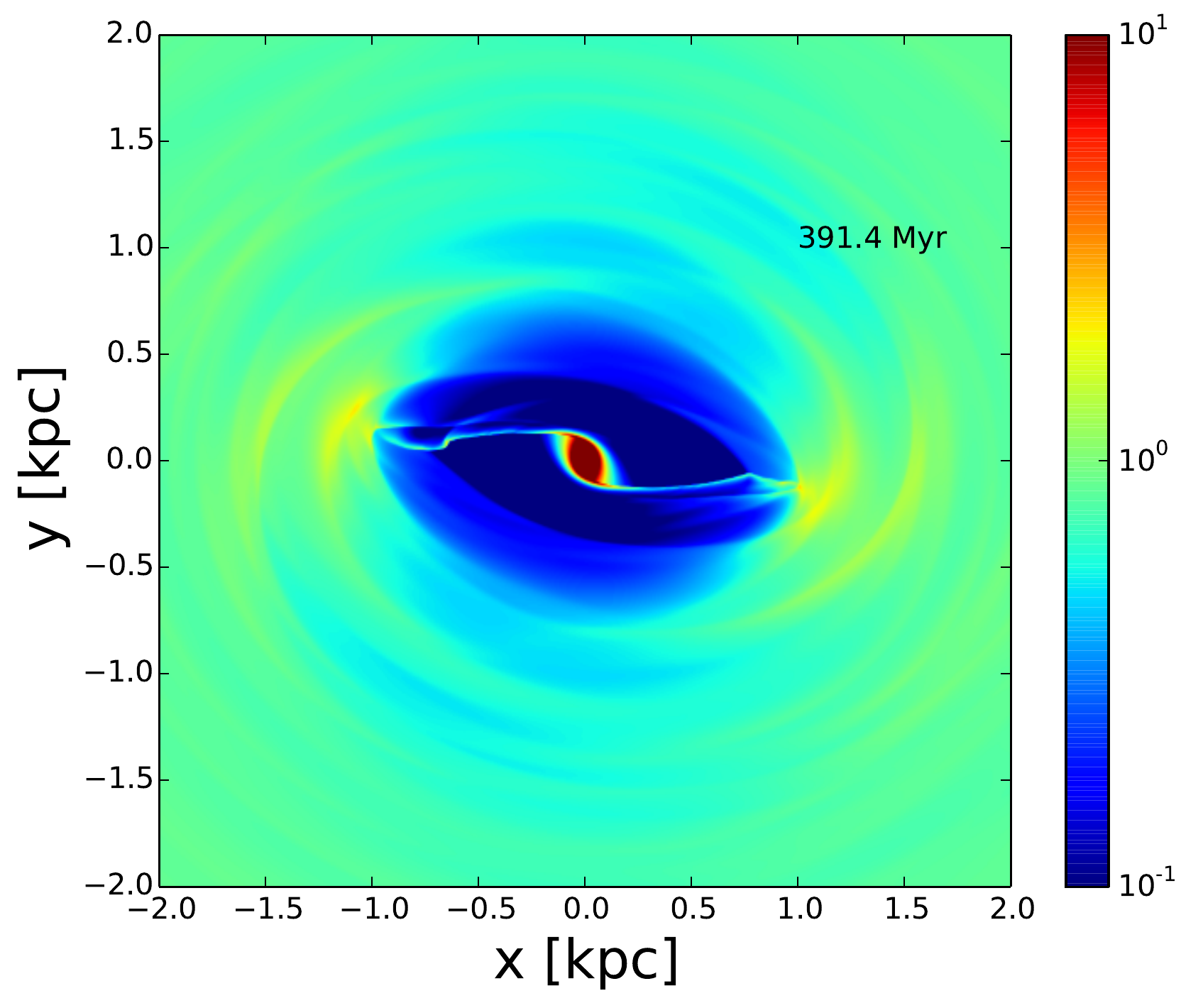}
\caption{Three late snapshots of a simulation using Pluto code and the MinMod limiter. This limiter allows the simulation to be only weakly unstable. It shows that shocks are almost formed and destroyed cyclically. Swirls produced at one shock can propagate and bump on the shock on the other side.}
\label{fig:instability2}
\end{figure}

Fig.  \ref{fig:instability2} shows the flow computed with the most diffusive
limiter (MINMOD) at three different times. It shows the shocks cyclically
straightening out and then developing wrinkles.  More unstable simulations
also display a cyclical tendency for the shocks to straighten out and
dissolve, but in these lower-diffusivity calculations the dissolution of the
shocks is more sudden. From these tests we conclude that unsteadiness is
not a feature of our particular hydro code, but is shared by other well tested
codes as well.

Turbulence is a consequence of high rates of shear in a high Reynolds number
flow. In the flows studied by engineers, for example in pipes and over
aircraft wings, turbulence arises in the thin boundary layer that forms when
a low-viscosity fluid meets a solid surface. Vortices formed in the boundary
layer move into the bulk of the fluid, making the flow generally unsteady. In
our simulations, vortices form along the shocks \citep[see][for an
analysis of vorticity generation in shocks]{Binney1974}, where the shear rate
is exceptionally high. The rate of shear is inversely proportional to the
width of the shock, so it increases with the grid resolution and decreases
with the diffusivity of the limiter.  Thus simulations with the finest grids
and the least diffusive limiters have the highest shear rates and are most
likely to become turbulent. 

The fact that  turbulence appears only in simulations with a high sound speed
can also be explained by the connection between shear rate and the onset of
turbulence: the rate of shearing increases along the sequence of $x_1$ orbits
as one approaches the cusped orbit, so at low sound speeds, when the shock
forms far from the cusped orbit, the maximum rate of shearing is smaller than
when the sound speed is larger.

\begin{figure*}
\includegraphics[width=1.0\textwidth]{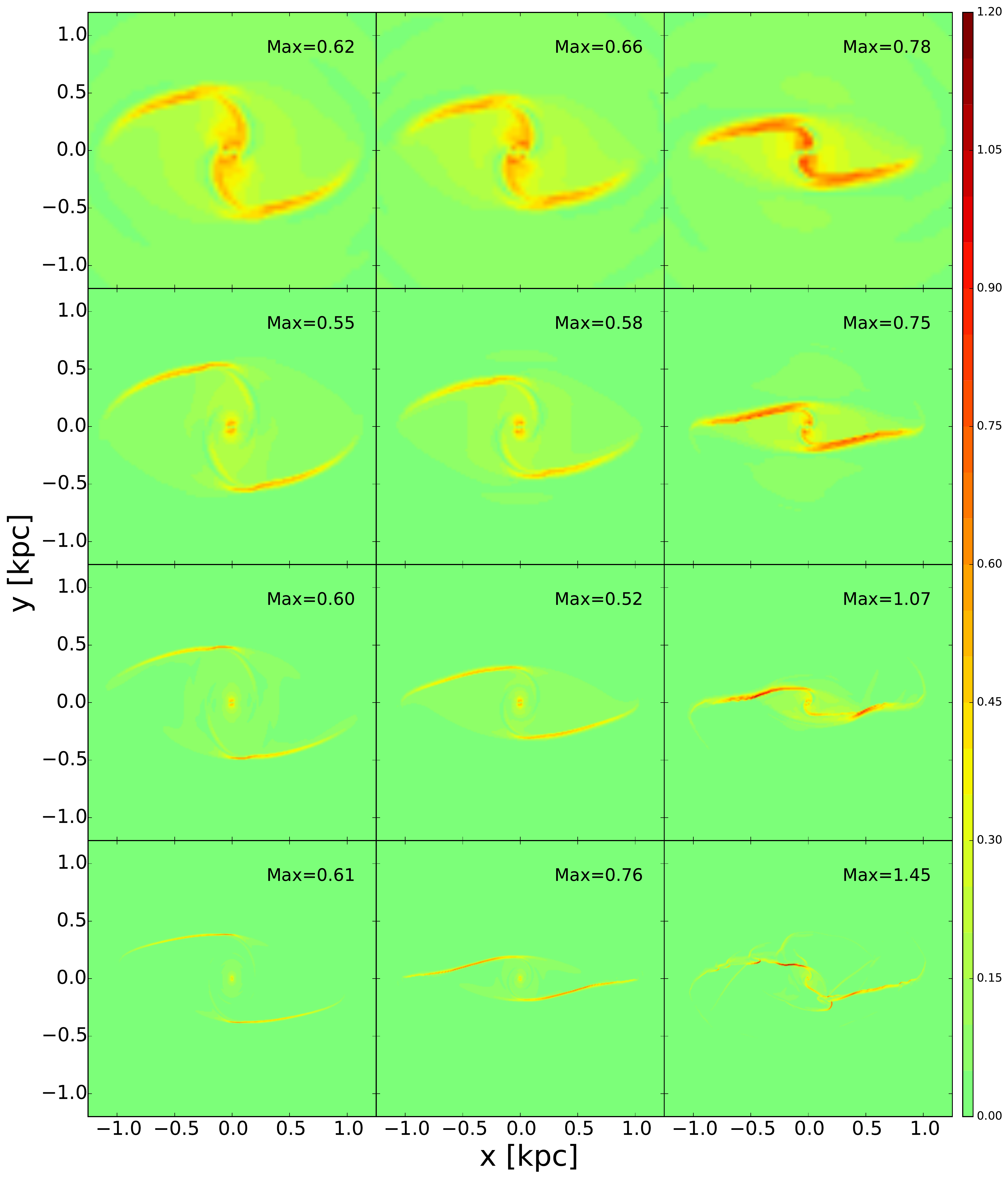}
 \caption{The quantity $\tau \di x$, where $\tau=$ is an indicator of the shear and is defined in Eq. \eqref{eq:shear}. In each panel, the maximum value reached by the quantity is shown.} \label{fig:shear}
\end{figure*}

A good indicator of the amount of shear in a 2D flow is the quantity \citep[see for example][]{Macie2008}
	\begin{equation} \tau^2 = \left(\frac{\pa v_x}{\pa y} + \frac{\pa v_y}{\pa x} \right)^2 + \left(\frac{\pa v_x}{\pa x} - \frac{\pa v_y}{\pa y} \right)^2.  \label{eq:shear} \end{equation}
This quantity is invariant under rotations of the coordinates, being the magnitude of the
eigenvalues of the traceless shear tensor, defined by
	\begin{equation} D_{ij} = \frac{1}{2} \left( \frac{\pa v_i}{\pa x_j} + \frac{\pa v_j}{\pa x_i} - \delta_{ij} ( \bmath{\nabla} \cdot \bfv) \right). \end{equation}
Fig.~\ref{fig:shear} shows the quantity $\tau \di x$, where the shear $\tau$ is estimated by finite differences. As claimed above,
the shearing rate is high along the shocks and is higher for higher sound speeds.

\cite{Kimetal2012} also encountered the onset of turbulence as isothermal gas
moves in a rotating barred potential, and our Fig \ref{fig:instability} is
similar to their Fig.4, panel (c). They point out that vortices generated in
one shock pass to the shock on the other side of the galaxy, and are there
amplified.  This process has been called the wiggle instability \citep[see
for example][and references
therein]{WadaKoda2004,KimKimKim2014}.  There is still no
consensus on the nature of this instability: \cite{WadaKoda2004} conjectured
it is a Kelvin-Helmholtz (KH) type instability, but \cite{KimKimKim2014} use
a shearing box analysis to argue it should be considered as a different type
of instability. The instability has also been attributed to a numerical noise
caused by the discretisation of the fluid equations \citep{HanawaKicuchi2012}.


\subsection{Numerical convergence}

We have seen that as we increase the resolution at given sound speed, the
shocks are postponed and the $x_2$ disc shrinks. Have we converged, or would a
further increase in resolution produce significant changes in the
flow? In order to discuss this, let us consider how the size of the $x_2$
disc depends on the resolution. In each snapshot in our simulations the
vertical edges of the $x_2$ disc can be clearly identified by a sharp
variation in the gas density as we move along the vertical axis.
Consequently,  the vertical size of the $x_2$ disc is well defined.

\begin{figure}
\includegraphics[width=0.5\textwidth]{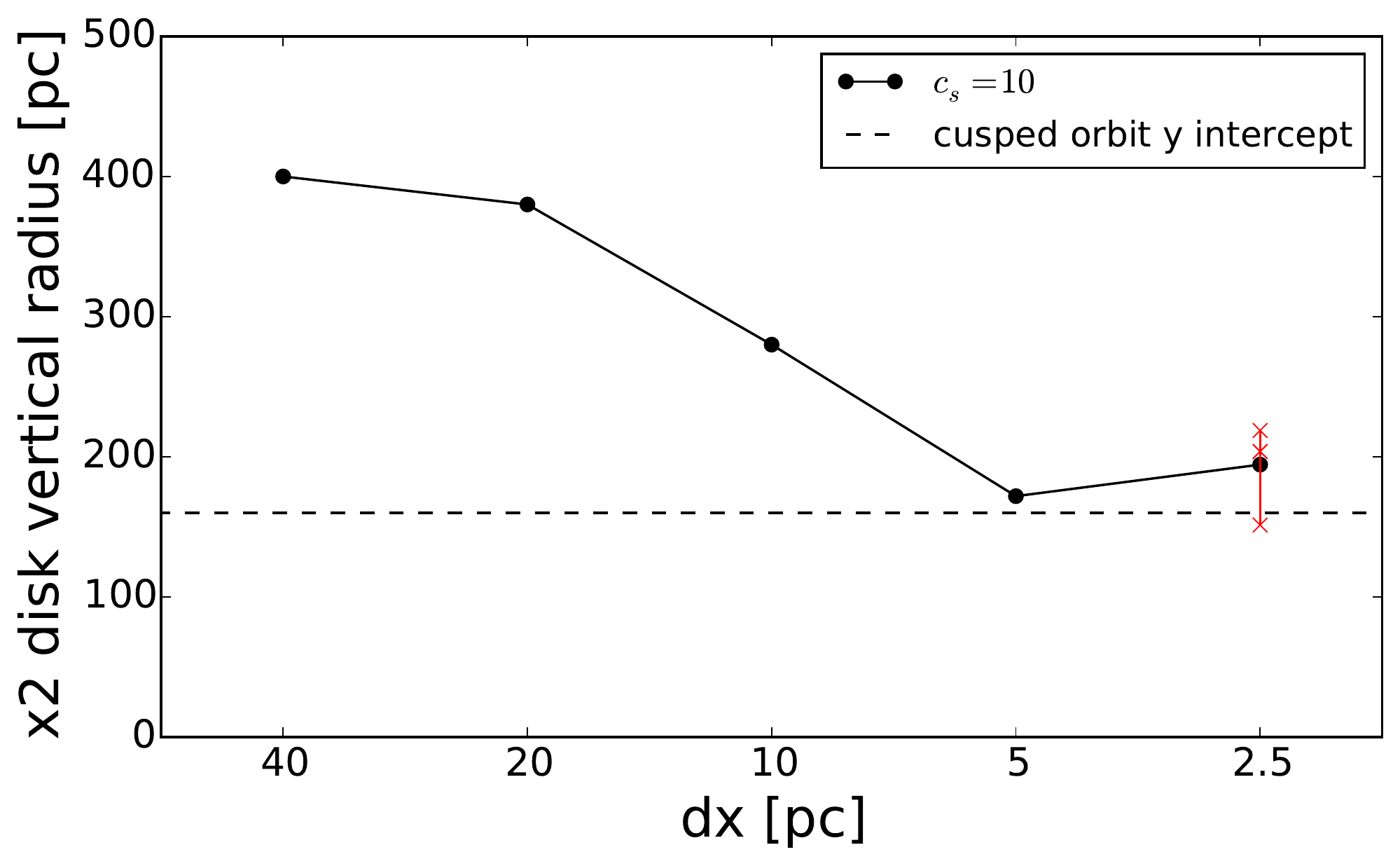}
\caption{The vertical radius of the $x_2$ disc as a function of resolution. Each point is obtained averaging the radius over 4 different snapshots at times $t=274,313,352,391$ Myr. For resolution $\di x = 2.5\pc$ the flow is unsteady and the size of the disc fluctuates, and the red crosses show the scatter in the size of the $x_2$ disc over the four snapshots. The horizontal dashed line shows the value of the intercept of the cusped orbit with the vertical axes.}
\label{fig:x2sizes}
\end{figure}

Fig.~\ref{fig:x2sizes} shows the vertical radius of the $x_2$ disc plotted
against resolution for simulations with sound speed $\cs=10\kms$. The data
points for $\di x\ge 5\pc$ are from the simulations presented above, while
the black dot at $\di x=2.5\pc$ is obtained for an additional simulation.
Since the computational cost of a simulation of given spatial extent rises as
$(\di x)^{-3}$, the additional simulation was run on a grid that covered an
area only $5\kpc \times 5\kpc$ in extent -- we have verified that the results of lower
resolution simulations for the region inside the shocks are unaffected by a
reduction of the extent of the grid, so the additional simulation should provide
a valid additional datum for the radius of the $x_2$ disc.

The simulation with $\di x=2.5\pc$ was unsteady as expected, given the
tendency for unsteadiness to arise at high resolutions and sound speeds.  On
account of the unsteadiness, the size of the $x_2$ disc in the $\di x=2.5\pc$
simulation fluctuates. In Fig.~\ref{fig:x2sizes}, the point for the $\di x =
2.5\pc$ simulation is obtained by averaging the $x_2$ disc size over four
different snapshots at times $t=274,313,352,391$ Myr, and the red crosses
show the values in each snapshot. Such averaging would not modify the data
points for the coarser, steady, simulations. Since in the simulation with
$\di x = 2.5\pc$ the mean radius of the $x_2$ disc is compatible with this
radius when $\di x= 5\pc$, we conclude that the size of
the $x_2$ disc has converged to near its true value at $\cs=10\kms$.

As can be seen from Fig. \ref{fig:x2sizes}, the value of the vertical radius
towards which we have converged for $\cs=10\kms$ is the value of the
intercept of the cusped orbit with the vertical axis. Inspection of
Fig.~\ref{fig:res1} shows that for higher sound speeds, the $x_2$ disc
becomes smaller. Indeed, \cite{englmaiergerhard1997} found that the $x_2$
disc disappears for high enough sound speed.

\subsection{A paradox} 

In the limit of vanishing pressure, the characteristics of the Euler
equations reduce to ballistic trajectories. Hence, we expect the ballistic
approximation to fluid flow to work best when the sound speed is low. Why is
the BGSBU picture, which is founded on the ballistic approximation, more
accurately reproduced with $\cs=10\kms$ than with $\cs=5\kms$?

The answer is that the BGSBU picture includes the requirement that the
transition between orbit families lies close to the cusped orbit.  This
transition is effected by shocks, which are more rather than less likely to
form in the limit of vanishing pressure. Shocks reflect the necessity in a
fluid for there to be a unique velocity at each spatial point. As
\cite{Riemann} showed, in certain circumstances the Euler equations predict
more than one velocity at a given point of a fluid in the sense that
characteristics of the same family can cross. The unphysical implications of
crossing characteristics are voided by a shock forming near the crossing
point. In the shock the finite mean-free path of the fluid's constituent
particles becomes important, and the Euler equations, which are based in the
fiction that the fluid forms a continuum, cease to be valid.

In a region of the disc where there are two distinct families of closed
orbits, the principle that in the limit of vanishing pressure the
streamlines/characteristics will be closed orbits does not suffice to
determine the flow. One needs in addition some way of selecting one orbit
family for the streamlines. Since the $x_1$ orbits do not exist at very small
radii and the $x_2$ orbits do not exist at very large radii, it is evident
that in a barred galaxy the flow must at some point transition from $x_1$ to
the $x_2$ orbits. Our simulations have shown that the location of the transition is controlled by the pressure 
even though the latter is too weak to be dynamically significant upstream of the transition. BGSBU
argued that the transition occurs at the cusped orbit, the latest that it
could occur, and with hindsight it is not surprising that this requires that
the sound speed is $\ga10\kms$.

\subsection{Sound speed and the nature of the ISM}
Our results can reproduce BGSBU predictions only for a particular sound speed.
Does this mean we are providing a measurement of the sound speed? 

In addressing this question, a major issue is that we have investigated only
one potential and only a single value of the pattern speed, so we cannot exclude the possibility that in another potential
the data could be fitted with a lower value of $\cs$. A trivial illustration
of this fact is provided by one of the axisymmetric potentials that were used
to interpret $(l,v)$ plots before BGSBU. However, the sharply peaking
circular-speed curves that these potentials require to fit the HI envelope in
the $(l,v)$ plane are highly implausible, and if the mechanism proposed by
BGSBU for generating this feature is to work, gas must stay close to $x_1$
orbits until rather close to the cusped orbit. That is, in a model with a
small value of $\cs$, the upward rise in the HI envelope would have to be at
least partly explained by inward-increasing values of the circular speed.
Moreover, the cusped orbit would have to occur at smaller radii than BGSBU
hypothesised.

Given the almost fractal nature of the ISM, the physical significance of the
parameter $\cs$ that plays such an important role in the simulations is far
from evident. It is, however, worth noting that we have pushed the resolution
so high that the cell size of our simulations is comparable to the size of a
giant molecular cloud -- even small GMCs are $\ga5\pc$ across.

\section{Implications  for the interpretation of observational data} \label{sec:discussion2}
\subsection{BGSBU revisited}
The three key points
in BGSBU's interpretation of the observational $(l,v)$ plots were:

\begin{enumerate}
\item The high velocities, reaching $v\simeq270\kms$, found in the HI $(l,v)$
data could be explained by gas moving on $x_1$ orbits just outside the
cusped orbit. When the $(l,v)$ plots had been interpreted in the context of
an axisymmetric Galaxy in previous studies \citep[see for
example][]{Sofue2013}, the high velocities observed had required the
circular-speed curve to first rise, and then fall with implausible rapidity.
By hypothesising a bar, BGSBU were able to explain the sharp peaks in the
$(l,v)$ envelope with a monotonically rising circular-speed curve. 

\item The parallelogram they identified in the CO emission was a cut version
of the longitude-velocity trace of the cusped orbit. The discrepancy between
the cusped orbit and the CO parallelogram was attributed to the limits of the
ballistic approximation, and it was hoped that full hydrodynamical
calculation would resolve this discrepancy. In particular, BGSBU pointed to
the hydrodynamical simulations of \cite{Athan92b}, which used the same
numerical method as this paper, to construct an qualitative argument that
qualitatively resolved the discrepancy.  

\item The emission at low longitudes visible in the CS was attributed to gas
on $x_2$ orbits. Outer $x_2$ orbits that can be found in the potential (see
Fig.  \ref{fig:orb1}) were assumed to be unoccupied by gas. 
\end{enumerate}
In the previous section we confirmed that the flow of gas can be understood
as a transfer from $x_1$ to $x_2$ orbits. However, the transition point is
not always at the cusped orbit, as BGSBU assumed, but depends on the
parameters of the gas flow, especially the sound speed. For the right sound
speed, $\cs=10\kms$, and our maximum resolution, $\di x=5\pc$, the gas flow
is very similar to that hypothesised by BGSBU. We now re-analyse the
observational data in light of this particular simulation.  Further below we
take a step back and discuss interpretation of data in a broader view. 

\begin{figure}
\centerline{
\includegraphics[width=0.4\textwidth]{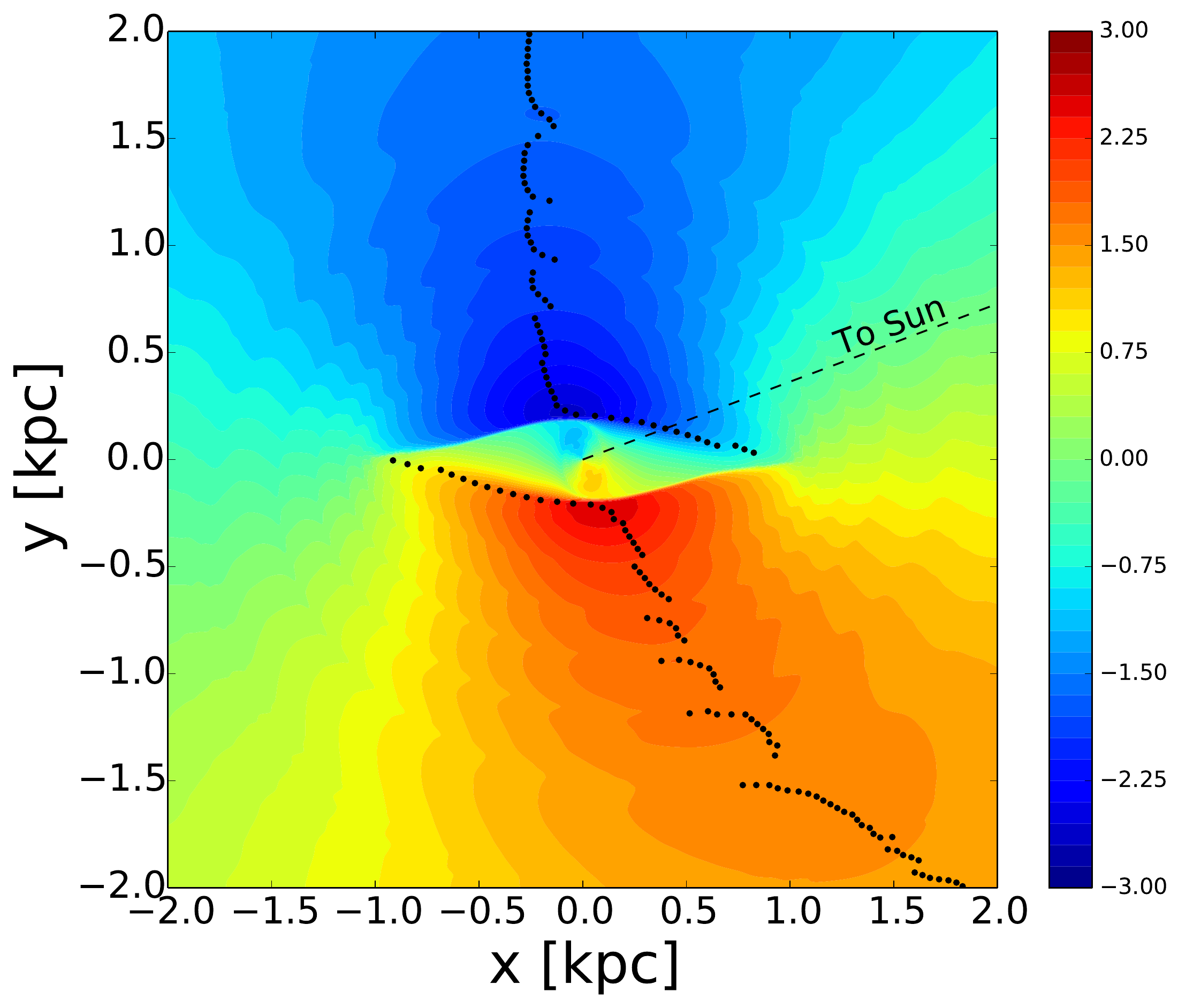}}
 \caption{The distribution of the projected line-of-sight velocity in the
$xy$ plane for the model $\di x=05\pc$, $\cs=10\kms$. The black dots show the
points corresponding to the envelope of gas in the $(l,v)$ plane. The angle
between the Sun-Galactic centre line and the bar major axis is assumed to be
$\phi=20\degree$. The colorbar is in units of $100\kms$.}
\label{fig:deprojection}
\end{figure}

In the centre-bottom panel of Fig.~\ref{fig:res3} we can see the $(l,v)$ plot
corresponding to the simulation with $\di x = 5\pc$ and $\cs = 10 \kms$. The
envelope of the hydro follows that of the closed orbits very well and
reproduces the high velocity peaks found in observations.
Fig.~\ref{fig:deprojection} shows how the line-of-sight velocity varies in
the $xy$ plane for the assumed observing angle $\phi=20\degree$. The black
dots show the locations that generate the envelope of emission in the $(l,v)$
plane. We can see that the high-velocity peaks at $|l|\simeq2\degree$ arise
from gas that lies very nearly along an $x_1$ orbit just outside the cusped orbit, as well as the portion of
envelope that runs from the peak down into the zone of forbidden velocities.  
Thus item (i) above is nicely confirmed by this hydro simulation. 

Well outside the cusped orbit  the black
dots in Fig.~\ref{fig:res2} have wiggles associated with spiral arms that run
out from the extremities of the bar.  These wiggles are reflected in small
oscillations in the $(l,v)$ envelope, qualitatively similar to what is
observed for the envelope of real observations. 

\begin{figure}
\includegraphics[width=0.4\textwidth]{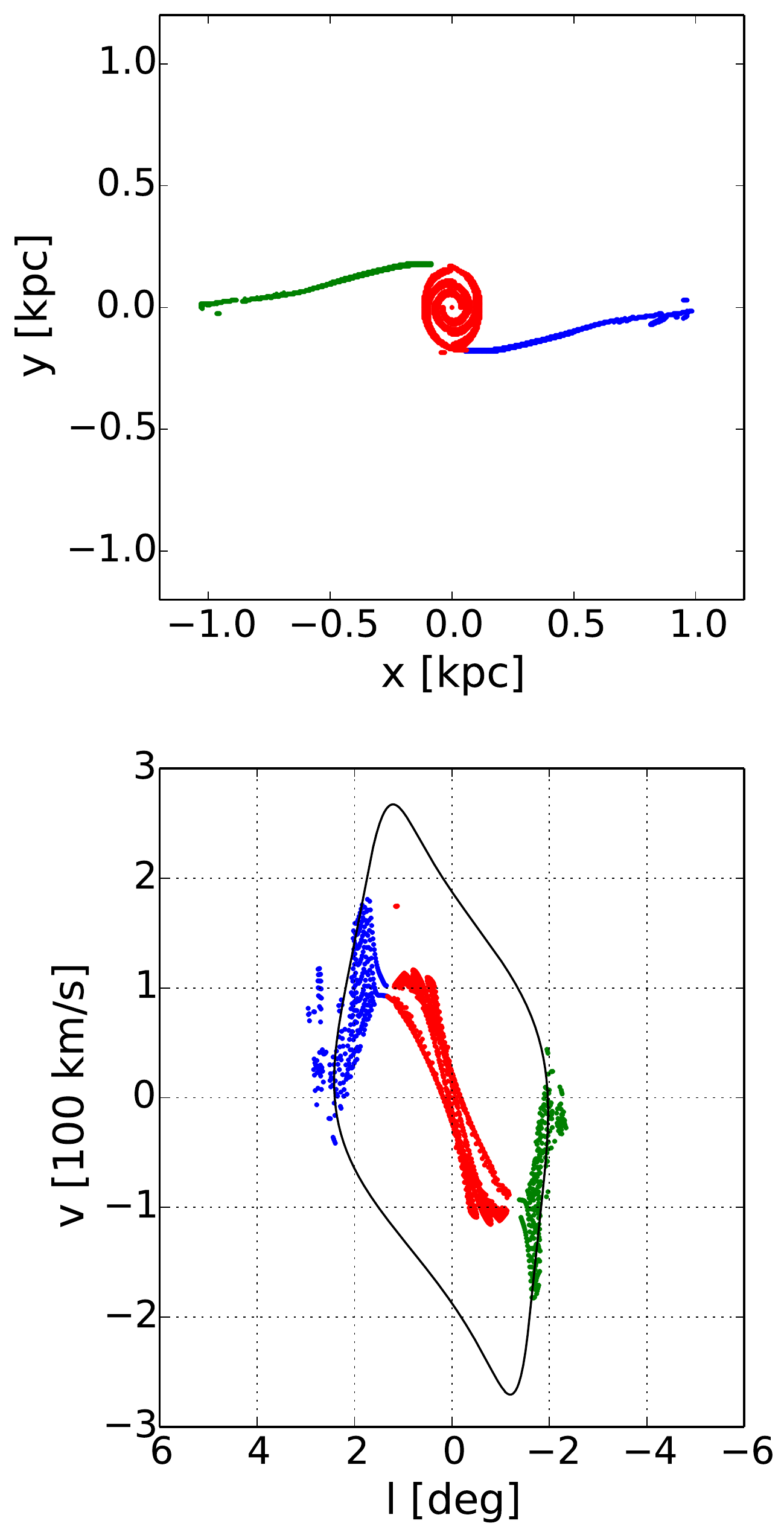}
 \caption{Top: points on the narrow shocks and the $x_2$ disc in the $xy$
plane. Bottom: their projected position in the $(l,v)$ plane. In black the
trace of the cusped orbit is shown.} \label{fig:patches}
\end{figure}

As mentioned above, two distinct structures resembling a parallelogram play a
role in the BGSBU picture: one is the CO parallelogram, presumed to be a cut
version of the cusped $x_1$ orbit, and a second is the trace of an $x_1$
orbit just outside the cusped orbit, which is responsible velocity peaks in
the envelope of HI emission. In the simulated $(l,v)$ plots we can identify
only one such parallelogram, namely the second one. How can we solve this inconsistency?
Fig.~\ref{fig:patches} shows the shocks (blue and green) and the $x_2$ disc
(red) in the $xy$ plane (top) and in the $(l,v)$ plane (below). In the lower
panel the shocks closely follow the vertical sides of the cusped orbit's
(black) parallelogram, but are cut almost exactly where the CO parallelogram
ends. We conclude that the vertical sides of the CO parallelogram are formed
by shocks. The other two sides of the CO parallelogram must be made up of gas
flowing from one shock to the other.  The shocks do not show up in our
$(l,v)$ projection of the gas flow, though they are quite apparent in the
$xy$ plane.  In reality, we expect the shocked gas to be brighter than our
simple minded radiative calculation suggests, because a lot of atomic gas is
converts to molecular as it is compressed at the shocks. 
When this effect is taken into account the second parallelogram should appear in the $(l,v)$ projection of the model. 
Thus we confirm item (ii) above of BGSBU's interpretation, although the physical mechanism that
causes the parallelogram to be cut was not exactly described by BGSBU.
Finally, Fig.~\ref{fig:patches} clearly shows that in the $(l,v)$ plane the
central molecular zone (red) occupies the region of the inner $x_2$ orbits,
as conjectured by BGSBU [item (iii) above].

When the $x_1/x_2$ transition happens well before the cusped orbit because
either the sound speed or the resolution is low, there is no gas able to
explain the high velocity peaks as in item (i). Moreover, the shocks project
to $|l|>2\degree$ so they cannot be associated with part of the CO
parallelogram, and the central molecular zone is predicted to extend to
higher $|l|$ than where significant CS emission is found, compromising the
interpretation of item (iii). Thus several different aspects of the data
point to gas remaining on $x_1$ orbits right up to the cusped orbit.


\subsection{The asymmetry}
It is well known that the molecular emission in the central molecular zone is
highly asymmetric: three quarters of the $^{13}$CO and CS emission comes from
positive longitudes. Perspective effects cannot account for this asymmetry
\citep{jenkinsbinney}, so the observed asymmetry must be transient:
observations made tens of megayears in the past or future would often show
asymmetry in the opposite sense. Thus  the likely
explanation of the asymmetry is unsteady flow, and the principal motivation
of the work reported in \cite{jenkinsbinney} was to find evidence of
unsteady flow. Although low-amplitude unsteadiness was generated in their
simulations, the present study implies that their simulations were too crude to
probe the physically interesting regime. 

We saw above that the greater prominence of the CO rectangle in the observed
$(l,v)$ plot than in the theoretical $(l,v)$ plot suggests that the shocks
are important sites for the conversion of atomic to molecular gas. This being
so, unsteady flow through the shocks will give rise to unsteady conversion of
atomic to molecular gas, so the atomic/molecular ratio on each side of the
Centre could well fluctuate as widely as the observations imply. However, a
full explanation of the observed asymmetries in molecular-gas emission must
await high-resolution simulations that keep track of the chemistry of the
ISM.

\subsection{What do we still need to explain in the $(l,v)$ diagram?}

Two  aspects of observations
remain inadequately  explained:
\begin{itemize}
 \item Coherent broad features like the $3\kpc$ arm and its counterpart on the
far side of the Galaxy \cite{Dame2008}. These are not produced in our
simulations. They {\it are\/} produced in other hydro simulations
\citep{mulderliem,combesrodriguez2008}, but these simulations did not
reproduce the high peaks in the envelope of HI emission in the $(l,v)$ plane,
probably for lack of resolution.
 \item Forbidden emission at large longitudes. The portion of the $(l,v)$
diagram covered by forbidden emission in our simulations is smaller than
the region in which coherent forbidden emission is seen in the data. A higher
quadrupole moment is probably needed to reproduce this. 
\end{itemize}
The essential elements of the BGSBU picture are that streamlines coincide with
$x_1$ and $x_2$ orbits, and that the shocks responsible for the
transition lie near the cusped orbit. BGSBU illustrated  these principles
with one particular, very simple potential. Better fits to the data could
surely be obtained with other, similar potentials. A fast way to select
potentials worthy of closer examination would be to use closed orbits
as BGSBU did.

\subsection{Relation to prior work} 

The question of what physical mechanism determines the size of the $x_2$ disc
is relevant for the interpretation of observations in our and external
galaxies \citep[see for example][]{Combes1996b,Kim2012b}. Our results suggest
that some previous studies may be biased by not taking into account the
effects of varying the resolution. For example, \cite{Cole2014} in their
simulations of galaxy formation found that the main mismatch between their
models and the observations was that the nuclear discs of their models were
too big relative to their bars. Since nuclear discs are $x_2$ discs, our
results suggest that the mismatch would be resolved by an increase in
resolution.

Finally, we mention that our findings do support the hopes of
\cite{bissantzetal2003}. These authors found, as in our low-resolution
simulations, that innermost non-self intersecting $x_1$ orbits near the
cusped orbit were unoccupied, and attributed this fact to details of the SPH
scheme, which do not apply for our grid-based simulations. This work gives
strong support to their view that in higher-resolution, grid-based simulations
their inner $x_1$ orbits would be occupied by gas, giving rise to peaks in their
$(l,v)$ projections.

\section{Conclusion} \label{sec:conclusion}
\cite{binneyetal1991} (BGSBU) constructed a picture of the flow of gas through
the central few kiloparsecs of our Galaxy. Their picture was based on the
idea that gas follows closed orbits, and it involved a particular choice of
orbit at which the gas transitions from the $x_1$ orbit family to the $x_2$
family. Their orbit-based picture required validation by hydrodynamical
simulations of gas flow. Early efforts in this direction did not provide the
necessary validation, in part because they did not adopt the same Galactic
potential as BGSBU, but largely because in them gas occupied some $x_2$
orbits that BGSBU required to be empty, and left empty some $x_1$ orbits that
BGSBU required to be occupied. We have run high-resolution, grid based hydro
simulations of gas flow in the potential of BGSBU and validated  their picture
in the case that the effective sound speed in the ISM is $\cs\ga10\kms$. 

The simulations confirm that, regardless of the sound speed adopted and the
grid resolution employed, gas streamlines closely coincide with closed orbits
everywhere outside a shock-dominated transition region that divides the outer
region, in which gas follows $x_1$ orbits, from an inner region, in which gas
follows $x_2$ orbits.  However, the orbit at which the shock arises, and the
transfer commences, depends on both the sound speed and the grid's
resolution.  Shock formation is favoured by both low sound speeds and low
grid resolutions, so increasing the sound speed and/or the grid resolution
moves inward the shock that causes gas to plunge from $x_1$ to $x_2$ orbits.
The BGSBU picture calls for the shock to occur as close to the Galactic
centre as it logically can, namely at the cusped orbit, interior to which
$x_1$ orbits become self-intersecting. Consequently, a flow consistent with
the BGSBU picture cannot be obtained with either a low sound speed or poor
spatial resolution.  It seems that previous simulations lacked the requisite
resolution. We find that a consistent flow can be obtained for $\cs\simeq10\kms$
and grid spacing $\d x\la5\pc$.

BGSBU did not provide a satisfactory explanation of the parallelogram-like
structure of CO emission in the $(l,v)$ plane. We find that the shocks form
two sides of the CO parallelogram and conjecture that the prominence of the
CO parallelogram is due to efficient conversion of
atomic gas into molecular gas. Unfortunately, we do not follow the ISM's
chemistry.

In our highest-resolution simulations the flow in the transition region
between the $x_1$ and $x_2$ orbits is unsteady. We think this unsteadiness is
probably a real physical phenomenon rather than a computational artifact, and
is essentially turbulence generated in the region of high shear behind the
shocks. We consider this unsteadiness, in conjunction with efficient
conversion of atomic gas to molecular form in the shocks, provides a
promising explanation of the observed asymmetry in the distribution of CO
emission either side of the Galactic centre.

While our simulations do provide strong support for the BGSBU picture, they
do not explain all aspects of the observed HI and CO emission. There is,
however, every prospect that further high-resolution simulations of flows in
potentials similar to that used by BGSBU will explain all significant
features. In this connection, two very worthwhile upgrades of our simulations
would be an increase in the quadrupole moment of the bar, and inclusion of
the conversion of gas between atomic and molecular forms.

\section*{Acknowledgements}

We are grateful to Witold Maciejewski for helpful comments. MCS acknowledges the support of the Clarendon Scholarship Fund and is
indebted to Steven N. Shore and Mir Abbas Jalali for helpful discussions.  
JB and JM were supported by Science and Technology Facilities Council by grants R22138/GA001 and
ST/K00106X/1.  JM acknowledges support from the ``Research in Paris''
programme of Ville de Paris.  The research leading to these results has received funding from
the European Research Council under the European Union's Seventh Framework
Programme (FP7/2007-2013) / ERC grant agreement no.\ 321067.

\def\aap{A\&A}\def\aj{AJ}\def\apj{ApJ}\def\mnras{MNRAS}\def\araa{ARA\&A}\def\aapr{Astronomy \&
  Astrophysics Review}\def\apjs{ApJS}
\bibliographystyle{mn2e}
\bibliography{SBM}

\end{document}